\documentclass[10pt,journal,final,letterpaper]{IEEEtran}

\usepackage{cite}

\usepackage[pdftex]{graphicx}
\usepackage{xcolor}

\usepackage{amsmath, amssymb, amsfonts}

\usepackage[ruled,norelsize]{algorithm2e}
\makeatletter
\newcommand{\removelatexerror}{\let\@latex@error\@gobble}
\makeatother

\usepackage{array}

\usepackage{mwe}
\ifCLASSOPTIONcompsoc
 \usepackage[caption=false,font=normalsize,labelfont=sf,textfont=sf]{subfig}
\else
 \usepackage[caption=false,font=footnotesize]{subfig}
\fi
\usepackage{url}
\usepackage{siunitx} 

\usepackage{lipsum}

\usepackage{todonotes} 
\definecolor{morange}{rgb}{0.8,0.2,0}
\definecolor{mblue}{rgb}{0,0.3,1.0}
\definecolor{mgreen}{rgb}{0.2,0.4,0}
\definecolor{mpurple}{rgb}{0.5 0.1 0.7}
\definecolor{mred}{rgb}{1,0,0}
\definecolor{cyan(process)}{rgb}{0.0, 0.72, 0.92}

\begin{document}

\title{Half-Space Modeling with Reflecting Surface in Molecular Communication}
%
%
%

\author{Anil Kamber,
        H. Birkan Yilmaz,~\IEEEmembership{Member,~IEEE,}
        Ali Emre Pusane,~\IEEEmembership{Senior Member,~IEEE,}
        Tuna Tugcu,~\IEEEmembership{Senior Member,~IEEE}
\thanks{A. Kamber and A. E. Pusane are with the Dept. of Electrical and Electronics Engineering, Bogazici University, Istanbul, Turkey. e-mail: anil.kamber@std.bogazici.edu.tr,\,ali.pusane@bogazici.edu.tr}
\thanks{H. B. Yilmaz and T. Tugcu are with NETLAB, Dept. of Computer Engineering, Bogazici University, Istanbul, Turkey. e-mail:\{birkan.yilmaz,\,tugcu\}@bogazici.edu.tr}
\thanks{}
}

%
%

\markboth{Submited to  IEEE TRANSACTIONS ON MOLECULAR, BIOLOGICAL, AND MULTI-SCALE COMMUNICATIONS}%
{Kamber \MakeLowercase{\textit{et al.}}: Half-Space Modeling with Reflecting Surface in Molecular Communication}



\maketitle

\begin{abstract}
In Molecular Communications via Diffusion (MCvD), messenger molecules are emitted by a transmitter and propagate randomly through the fluidic environment. In biological systems, the environment can be considered a bounded space, surrounded by various structures such as tissues and organs. The propagation of molecules is affected by these structures, which reflect the molecules upon collision. Deriving the channel response of MCvD systems with an absorbing spherical receiver requires solving the 3-D diffusion equation in the presence of reflecting and absorbing boundary conditions, which is extremely challenging. In this paper, the method of images is brought to molecular communication (MC) realm to find a closed-form solution to the channel response of a single-input single-output (SISO) system near an infinite reflecting surface. We showed that a molecular SISO system in a 3-D half-space with an infinite reflecting surface could be approximated as a molecular single-input multiple-output (SIMO) system in a 3-D space, which consists of two symmetrically located, with respect to the reflecting surface, identical absorbing spherical receivers.
\end{abstract}

\begin{IEEEkeywords}
Molecular communications, nanonetworking, diffusion channel, half-space, bounded environment, reflective boundary.
\end{IEEEkeywords}

\IEEEpeerreviewmaketitle

\section{Introduction}
\IEEEPARstart{M}{olecular} communication (MC) is a novel communication technology. In particular, MCvD has been gaining popularity as a promising approach \cite{kuran2020survey,gursoy2022towards}. MCvD systems could be utilized in complex health applications such as health monitoring, tissue engineering, and environment monitoring \cite{NAKANO,IAN,bi2021survey,ningthoujam2021implementing}, owing to their power-efficient nature \cite{energy,chouhan2023interfacing}. In MCvD systems, the message is transmitted through information-carrying molecules that propagate in the diffusive fluidic medium. Hence, in the absence of any flow in the environment, the motion of molecules is solely governed by Brownian motion. A typical MCvD system consists of either a single or multiple transmitters communicating with a single or multiple receivers.

In the context of communication theory, it is important to model communication systems analytically. For MCvD systems, distribution models for received signals are investigated in \cite{receivedsignaling}. Additionally, the mathematical formulation of MCvD systems with an absorbing spherical receiver in 3-D space is well-studied in the literature \cite{tunabirkan}. However, 3-D unbounded space is not an accurate approximation for biological environments. For instance, the impulse response of the
diffusion channel with a spherical absorbing receiver within a spherical
reflective boundary is presented in \cite{fatihdinc} to model the intra-cellular diffusion environment. Moreover, a spherical absorbing boundary is considered in \cite{absorbingspherical}. Many other approaches that incorporate the effect of boundary \cite{synapticcell,vessel,partially,chip,bloodvess} could be listed. 

Nanonetworks, which operate in the vicinity of biological structures such as tissues and organs for the purpose of tissue engineering or environment monitoring, would have been affected by the reflections caused by these structures. If this effect is not compensated for or anticipated beforehand, the quality of communication significantly drops, leading to loss of information. Moreover, with advancements in MCvD systems, more crowded scenarios that include multiple receivers operating near biological structures might be feasible. For this purpose, it is important to derive a mathematical formulation of the channel response of MCvD systems near a reflecting surface. 

Derivation of the channel response in 3-D half-space requires solving the diffusion equation in 3-D space, which is a partial differential equation system with one initial and three boundary conditions. A general solution for the diffusion equation in 3-D half-space with a reflecting surface is proposed in \cite{schulten2000lectures}. However, to derive the channel response of an MCvD system with an absorbing spherical receiver, consideration of the spherical absorbing boundary is needed in addition to the reflecting boundary condition. The procedure used to derive the general solution becomes intractable when applied to deriving the channel response. On the other hand, the procedure followed in \cite{tunabirkan} to derive the channel response becomes intractable in the presence of a reflecting boundary condition. Hence, a new approach, which could possess more physical intuition, is needed to derive the channel response in 3-D half-space. 

In this paper, the method of images is introduced to the MC realm, and our main aim is to present a new method for modeling 3-D half-space MCvD channels by utilizing the implications of Brownian motion and the nature of reflecting surfaces. To solidify the proposed model, the performance of the channel transfer function is tested with different topologies. After that, a special case where an MCvD system is confined to a bounded region by two parallel reflecting surfaces.

\begin{figure}[t]
\centering
\includegraphics[width = 0.7\columnwidth]{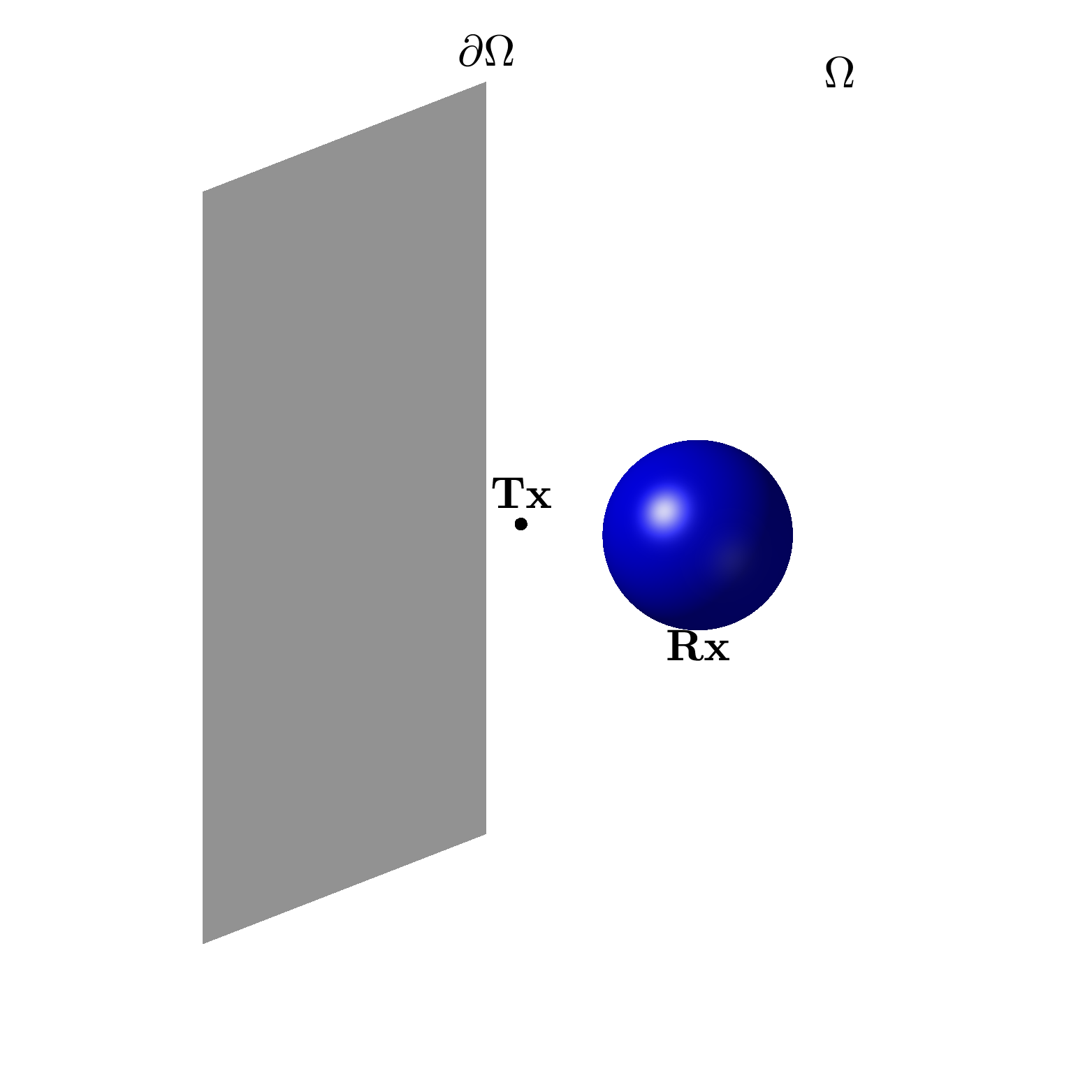}
\caption{\label{fig: MCvD_Basic_Drawing_Fig} Half-space MCvD system with an infinite reflecting surface.}
\end{figure}

The organization of this paper is as follows: Section II presents the molecular SISO topology in a half-space with an infinite reflecting surface. Subsequently, Section III-A derives and presents the analytical expression for half-space modeling with an infinite reflecting surface. In Subsection III-B, the derivation of the channel response in a 3-D half-space with an infinite reflective surface is provided. Following this, Subsection III-C examines a special case of half-space modeling, where an MCvD system is bounded by two parallel infinite reflecting surfaces, and provides an analytical expression. Section IV evaluates the accuracy of these models, and Section V concludes the paper.

\section{System Model}
\label{System Model Sec}

In this work, diffusion-based communication systems with an absorbing spherical receiver is considered in a 3-D half-space with an infinite reflecting surface, as shown in Fig. \ref{fig: MCvD_Basic_Drawing_Fig}. The environment is filled with fluid, and we assume that there is no drift. The molecular channel is in a 3-D unbounded space that is interrupted by an infinite reflecting surface denoted as $\partial \Omega$, and the channel is represented by Cartesian coordinate system. $\Omega$ represents the half-space, and $\partial \Omega$ represents the set of points in $\mathbb{R}^3$ that constitute the reflecting surface. For simplicity, assume that $\partial \Omega$ lies on the $yz$-plane, leading to
\begin{equation}
    \forall \Vec{P} = (x,y,z) \in \partial \Omega \Longleftrightarrow x = 0.
\end{equation}
Rx denotes a fully absorbing spherical receiver whose center is located at an arbitrary point $\Vec{C}$ = $(x_{R},y_{R},z_{R})$, where $x_{R} > 0$, and $r_r$ denotes its radius. Tx denotes an infinitesimal point transmitter that transmits molecules once, as an impulse, at a particular point, $t = 0$. It is located at $\Vec{T_x}$ = $(x_{T},y_{T},z_{T})$, where $x_{T} > 0$, and the distance between $\Vec{C}$ and $\Vec{T_x}$ is $r_0$, where $r_0 > r_r$. The channel is solely governed by Brownian motion.

For instance, an infected tissue is wanted to be monitored through using Rx. Molecules are disseminated from the infection site, which can be conceptualized as Tx. Molecules emanating from the site of infection are likely to be reflected by adjacent histological structures. Given the substantial size disparity between these histological structures and the Rx, the structures can be effectively modeled as an infinite reflecting surface.

To understand the effect of $\partial \Omega$ on the channel and its implications for the solution, let's first omit $\partial \Omega$. In the absence of $\partial \Omega$, the hitting rate of the molecules to Rx in a molecular SISO system $p_{\text{hit}}^{\text{SISO}}(\text{Rx}, t \mid r_0, r_r)$ is derived in \cite{tunabirkan} as

\begin{equation}
\begin{split}
    p_{\text{hit}}^{\text{SISO}}(\text{Rx}, t \mid r_0, r_r) & = \frac{r_r}{r_0} \frac{1}{\sqrt{4\pi Dt}} \\
    &\times \frac{r_{0}-r_{r}}{t} \exp\left(-\frac{(r_0 - r_r)^2}{4Dt}\right),
\end{split}
\end{equation}
where $D$ is the diffusion coefficient. 

Derivation of (2) could be developed from the solution of Fick's 3-D diffusion equation, which is presented in \cite{tunabirkan}, \cite{schulten2000lectures}. Since any kind of distortion that does not permit spherical symmetry in 3-D space is absent here, the spherical coordinate space could be utilized to solve
\begin{equation}
    \frac{\partial p(r, t \mid r_0)}{\partial t} = D \nabla^2 p(r, t \mid r_0),
\end{equation}
under certain initial and boundary conditions. The initial condition is defined as 
\begin{equation}
    p(r, t \rightarrow 0\mid r_0) = \frac{1}{4\pi r_0^2} \delta(r - r_0).
\end{equation}
This condition refers to the impulse function that is given to the system at $t = 0$ in order to derive the impulse response of the system, $p_{\text{hit}}^{\text{SISO}}(\text{Rx}, t \mid r_0, r_r)$, which is also defined as the channel transfer function. The first boundary condition is 
\begin{equation}
    \lim_{r \to \infty} p(r, t \mid r_0) = 0,
\end{equation}
where it can be assumed that the distribution vanishes at distances far from Rx. The second boundary condition is
\begin{equation}
    D \frac{\partial p(r, t \mid r_0)}{\partial r} = w p(r, t \mid r_0) \text{ for } r = r_r,
\end{equation}
where $w$ is the rate of reaction. When the spherical boundary is fully absorbing, $w \rightarrow \infty$, the distribution vanishes in the vicinity of the spherical receiver. This implies that the molecules are absorbed and removed from the system.

Now assume that the 3-D space is interrupted by $\partial \Omega$. To solve Fick's 3-D diffusion equation, another boundary condition is needed, which is defined as
\begin{equation}
    \Vec{a}(r) \cdot \Vec{\boldsymbol{J}}p(r,t|r_{0}, t = 0) = 0 \text{ } \forall r \in \partial \Omega,
\end{equation}
where $\Vec{a}(r)$ denotes the unit normal vector of the reflecting surface at $r$, and $\Vec{\boldsymbol{J}}p(r,t|r_{0}, t = 0)$ is the flux of molecules. This expression could be interpreted as
\begin{equation}
    \Vec{\boldsymbol{J}}p(r,t|r_{0}, t = 0) = D \nabla p(r, t \mid r_0, t = 0),
\end{equation}
which is also known as Fick's First Law of Diffusion.

At first, it is noticed that
\begin{IEEEeqnarray}{rCl}
\nabla^2 &=& \frac{1}{r^2} \left[ \frac{\partial}{\partial r} \left( r^2 \frac{\partial}{\partial r} \right) \right. \left. +\> \frac{1}{\sin^2 \theta} \frac{\partial^2}{\partial \phi^2} \right. \nonumber \\
&& \left. +\> \frac{1}{\sin \theta} \frac{\partial}{\partial \theta} \left( \sin \theta \frac{\partial}{\partial \theta} \right) \right].
\end{IEEEeqnarray}
Due to the presence of the reflective boundary condition (7), $p(r,t|r_{0}, t = 0)$ is not spherically symmetric. Therefore, the terms with $\theta$ and $\phi$ cannot be omitted as was done in \cite{tunabirkan}. As a result, the mathematical procedures and methods that are presented in \cite{tunabirkan} to solve Fick's 3-D diffusion equation become intractable in the presence of the reflecting boundary condition. Even though the general solution of the diffusion equation is already proposed in \cite{schulten2000lectures}, derivation of the absorption rate of molecules in the vicinity of Rx is very complicated under the second boundary condition (6). Hence, an appropriate modeling that requires more physical intuition is needed to approximate the solution of (3) under initial and boundary conditions (4), (5), (6), and (7), respectively.  

\section{Methodology}
\subsection{Method of Images}
Brownian motion is a Wiener process, in which each incremental step is independent of the previous steps. In the vicinity of $\partial \Omega$, the Wiener process is defined as
\begin{IEEEeqnarray}{rCl}
x(t + \Delta t) &=& 
\begin{cases} 
x'(t) , & \text{if } x'(t)  \in \Omega, \\
R_{\partial \Omega} \left[x'(t)  \right], & \text{if } x'(t) \notin \Omega,
\end{cases} \nonumber \\
&&
\end{IEEEeqnarray}
where $x'(t) = x(t) + \sqrt{2D \Delta t} \,  r(t)$ and $x(t)$ are the position vectors of a molecule in $\Omega$. $r(t)$ is the random unit vector that indicates the direction of the incremental step, and $R_{\partial \Omega}$ can be denoted as the operation of reflection:
\begin{figure}[t]
\centering
\includegraphics[width=0.7\columnwidth,keepaspectratio]{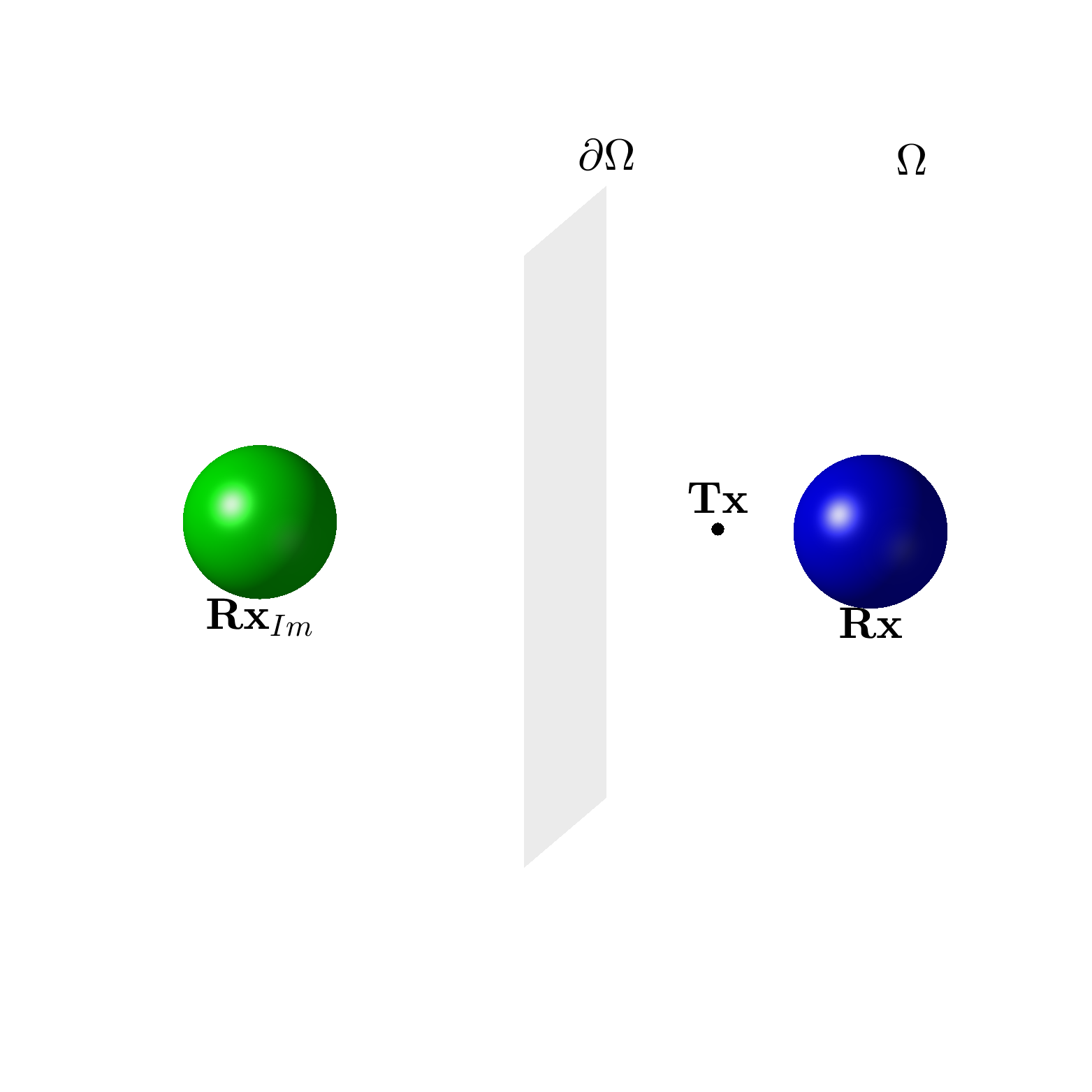}
\caption{\label{fig: Half-Space} Approximated SIMO system when the infinite reflecting surface is omitted.}
\end{figure}

\begin{equation}
\begin{aligned}
x'_{\parallel} & \xrightarrow{R_{\partial \Omega}} x'_{\parallel},\\
x'_{\perp} & \xrightarrow{R_{\partial \Omega}} 2\omega_{\perp} - x'_{\perp} ,
\end{aligned}
\end{equation}
where $\omega \in \partial \Omega$. $x'_{\parallel}$ denotes the component of $x'(t)$ that is parallel to $\partial \Omega$, and $x'_{\perp}$ denotes the component of $x'(t)$ perpendicular to $\partial \Omega$ \cite{schulten2000lectures}. The physical interpretation of (11) is the reflection of a particle in the direction normal to $\partial \Omega$. From the implications of (7) and (8), the following proposition could be deduced: 

\textbf{Proposition 1:} \textit{After reflection, the molecule continues along the trajectory of its mirror image.}

Let's define the absorption rate of molecules to Rx in $\Omega$ as $p_{\text{hit}}^{\text{SISO}}(\text{Rx}, t \mid r_0, r_r,\partial \Omega)$. Without loss of generality, $ p_{\text{hit}_{n}}^{\text{SISO}}(\text{Rx}, t \mid r_0, r_r,\partial \Omega)$ can be defined as the probability of being absorbed by Rx after being reflected $n$ times by $\partial \Omega$, where $n \in \mathbb{N}$. This implies
\begin{equation}
    \begin{split}
    p_{\text{hit}}^{\text{SISO}}(\text{Rx}, t \!\mid\! r_0, r_r, \partial \Omega) \!= & \ p_{\text{hit}_{0}}^{\text{SISO}}(\text{Rx}, t \mid r_0, r_r, \partial \Omega)  \\
    & + \ \ldots  \\
    & + \ p_{\text{hit}_{n}}^{\text{SISO}}(\text{Rx}, t \mid r_0, r_r, \partial \Omega),
    \end{split}
\end{equation}
when $n \rightarrow \infty$. From now on, $ p_{\text{hit}_{n}}^{\text{SISO}}(\text{Rx}, t \mid r_0, r_r,\partial \Omega)$ can be denoted as $p_{n}^{\text{H-SISO}}$ for simplicity. Hence,
\begin{equation}
    p_{n}^{\text{H-SISO}} = p_{0}^{\text{H-SISO}}+p_{1}^{\text{H-SISO}}+p_{2}^{\text{H-SISO}}+...+ p_{\infty}^{\text{H-SISO}}.
\end{equation}

Since Brownian motion is characterized as a Wiener process, any incremental step taken after $t_r$ seconds after the reflection event is independent of that event, particularly as $t_r \rightarrow 0$. Hence, with this understanding of Wiener processes and applying Proposition 1, the following could be deduced:

\textbf{Corollary 1:} \textit{When the reflecting surface is omitted, the sum of a molecule's probability of being absorbed by Rx and the molecule's probability of crossing the boundary of the mirror image of Rx, which is symmetric with respect to $\partial \Omega$, is equal to  $p_{\text{hit}}^{\text{SISO}}(\text{Rx}, t \mid r_0, r_r, \partial \Omega)$.}

(13) can be rearranged as
\begin{equation}
    \begin{split}
    p_{n}^{\text{H-SISO}} & = p_{0}^{\text{H-SISO}}+p_{2}^{\text{H-SISO}}+p_{4}^{\text{H-SISO}}+...+ p_{2n}^{\text{H-SISO}} \\
     & +  p_{1}^{\text{H-SISO}}+p_{3}^{\text{H-SISO}}+p_{5}^{\text{H-SISO}}+...+ p_{2n+1}^{\text{H-SISO}},
    \end{split}
\end{equation}
when $n \rightarrow \infty$. Crossing the boundary of the image of Rx, with respect to $\partial \Omega$, is tantamount to being absorbed by an imaginary receiver Rx$_{\text{Im}}$ that is identical to Rx as shown in Fig. \ref{fig: Half-Space}. Being reflected an odd number of times is equivalent to transition from real space to imaginary space when $\partial \Omega$ is omitted. Therefore,

\begin{equation}
\begin{split}
    p_{\text{hit}}^{\text{SIMO}}(\text{Rx}, t \mid \text{Rx$_{\text{Im}}$},r_0, r_r) = p_{0}^{\text{H-SISO}}+p_{2}^{\text{H-SISO}}+...+ p_{2n}^{\text{H-SISO}}, \\
    p_{\text{hit}}^{\text{SIMO}}(\text{Rx$_{\text{Im}}$}, t \mid \text{Rx},\text{$r_{\text{Im}}$}, r_r) = p_{1}^{\text{H-SISO}}+p_{3}^{\text{H-SISO}}+...+ p_{2n+1}^{\text{H-SISO}},
\end{split}
\end{equation}
when $n \rightarrow \infty$, and \text{$r_{\text{Im}}$} denotes the center-to-center distance between Tx and \text{Rx$_{\text{Im}}$}. As a result,
\begin{equation}
    \begin{split}
    p_{\text{hit}}^{\text{SISO}}(\text{Rx}, t \mid r_0, r_r, \partial \Omega) &=  \ p_{\text{hit}}^{\text{SIMO}}(\text{Rx}, t \mid \text{Rx$_{\text{Im}}$},r_0, r_r)  \\
    & + p_{\text{hit}}^{\text{SIMO}}(\text{Rx$_{\text{Im}}$}, t \mid \text{Rx},\text{$r_{\text{Im}}$}, r_r).
    \end{split}
\end{equation}

\textbf{Corollary 2:} \textit{A molecular SISO system in a 3-D half-space with an infinite reflecting surface could be approximated as a molecular SIMO system in a 3-D space, which consists of two symmetrically located, with respect to the reflecting surface, identical absorbing spherical receivers.}

\subsection{Channel Response}
A SIMO system with two identical Rxs is already investigated in \cite{Gokberk}. The absorption rate of molecules to each receiver in a SIMO system, $p_{\text{hit}}^{\text{SIMO}}(\text{Rx}_{i}, t \mid \text{Rx}_{j})$, is derived as
\begin{figure}[ht]
\centering
\includegraphics[width=0.7\columnwidth,keepaspectratio]{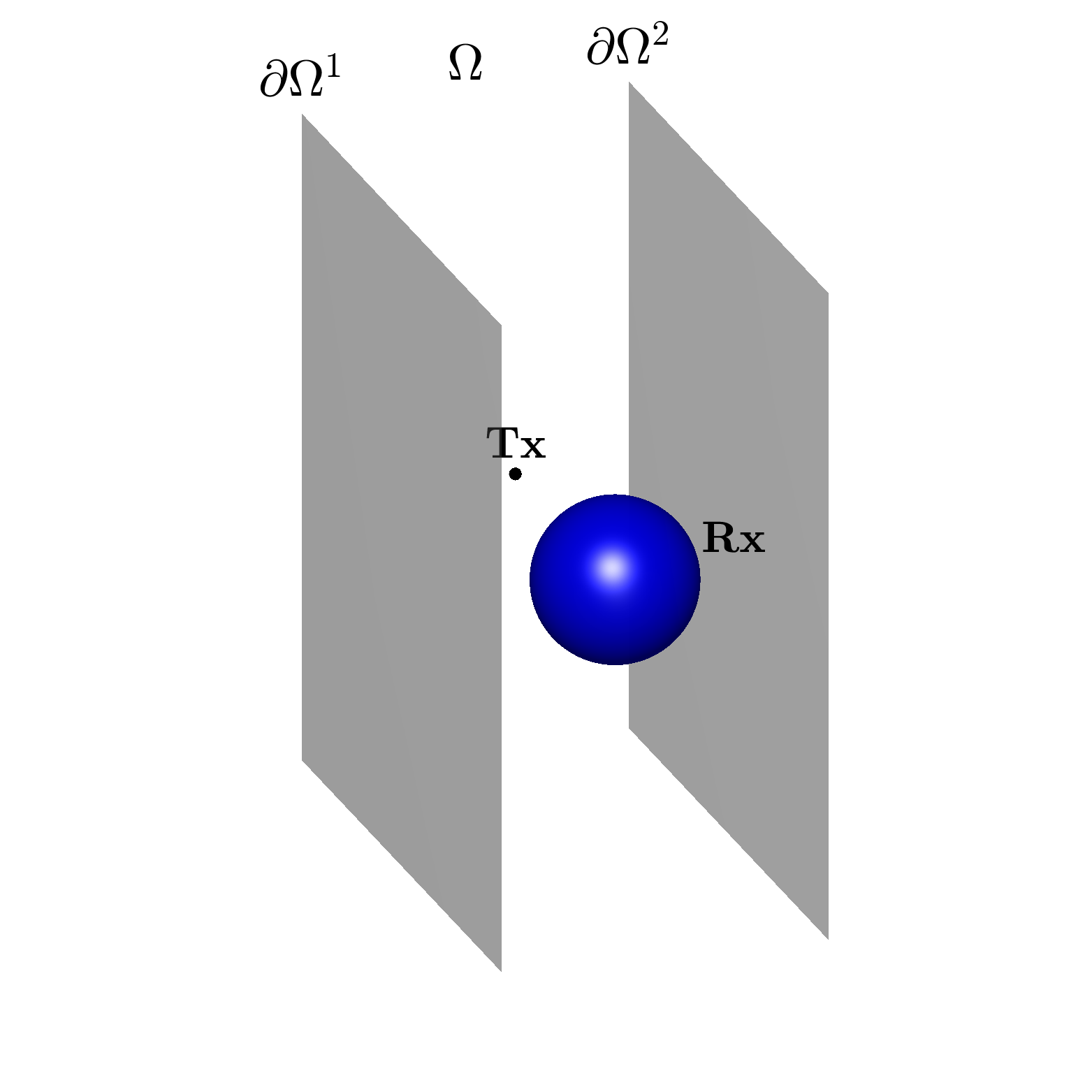}
\caption{\label{fig:TwoSurfaces} MCvD system between two infinite reflecting surfaces.}
\end{figure}

\begin{align}
 \hat{p}_{\text{hit}}^{\text{SIMO}}(\text{Rx}_{i}, t \mid \text{Rx}_{j}) &\!=\! \frac{r_i}{{r_0}_i} \frac{1}{\sqrt{4\pi Dt}} \frac{({r_0}_i - r_{i})}{t} \nonumber  \\
 &\!\times\! \exp\left(-\frac{({r_0}_i - r_i)^2}{4Dt}\right)  \\
   &\!-\! \frac{r_j r_i}{{r_0}_j {r_0}_{i|j}} \frac{1}{\sqrt{4\pi Dt}} \frac{({{r_0}_j} + {r_0}_{i|j}) - (r_j + r_i)}{t} \nonumber \\                
   &\!\times\! \exp\left(-\frac{({r_0}_j + {r_0}_{i|j}) - (r_j + r_i)^2}{4Dt}\right) \nonumber
\end{align}
in \cite{Gokberk}, and a more computationally feasible closed-form solution for the absorption probability of molecules to the $\text{Rx}_i$ in the SIMO system is derived through neglecting the absorption rate terms of recursively affected molecules as

\begin{equation}
p_{\text{hit}}^{\text{SIMO}}(\text{Rx}_{i}, t \mid \text{Rx}_{j}) = \int_{0}^{t} \hat{p}_{\text{hit}}^{\text{SIMO}}(\text{Rx}_i, \tau \mid \text{Rx}_j) d\tau \nonumber
\end{equation}
\begin{equation}
\begin{aligned}
&= \frac{r_i}{{r_0}_i} \, \text{erfc}\left( \frac{{r_0}_i - r_i} {\sqrt{4Dt}} \right)  \\
&- \frac{r_j r_i}{{r_0}_j {r_0}_{i|j}} \, \text{erfc}\left(- \frac{({r_0}_i + {r_0}_{i|j}) - (r_i + r_j)}{\sqrt{4Dt}} \right),
\end{aligned}
\end{equation}
where $r_i$ and $r_j$ are the radii of the receivers $\text{Rx}_{i}$ and $\text{Rx}_{j}$, respectively. $i$ and $j$ denote the indices of receivers. ${r_0}_i$ is the center-to-center distance between Tx and $\text{Rx}_{i}$, and ${r_0}_j$ is the center-to-center distance between Tx and $\text{Rx}_{j}$. ${r_0}_{i|j}$ and ${r_0}_{j|i}$ are then calculated by
\begin{equation}
    {r_0}_{i|j} = \sqrt{\left({r_0}_j -  {r_j}\frac{r_j}{{r_0}_j }\right)^2 + {{r_0}^2_i} - 2 \left({r_0}_j -  {r_j}\frac{r_j}{{r_0}_j }\right) {r_0}_i \cos \phi},
\end{equation}

\begin{equation}
    {r_0}_{j|i} = \sqrt{\left({r_0}_i -  {r_i}\frac{r_i}{{r_0}_i }\right)^2 + {{r_0}^2_j} - 2 \left({r_0}_i -  {r_i}\frac{r_i}{{r_0}_i }\right) {r_0}_j \cos \phi},
\end{equation}
where $\phi$ is the topology parameter that denotes the angular separation between $\text{Rx}_i$ and $\text{Rx}_{j}$ as stated in \cite{Gokberk}.

\begin{figure}[ht]
\centering
\includegraphics[width=0.7\columnwidth]{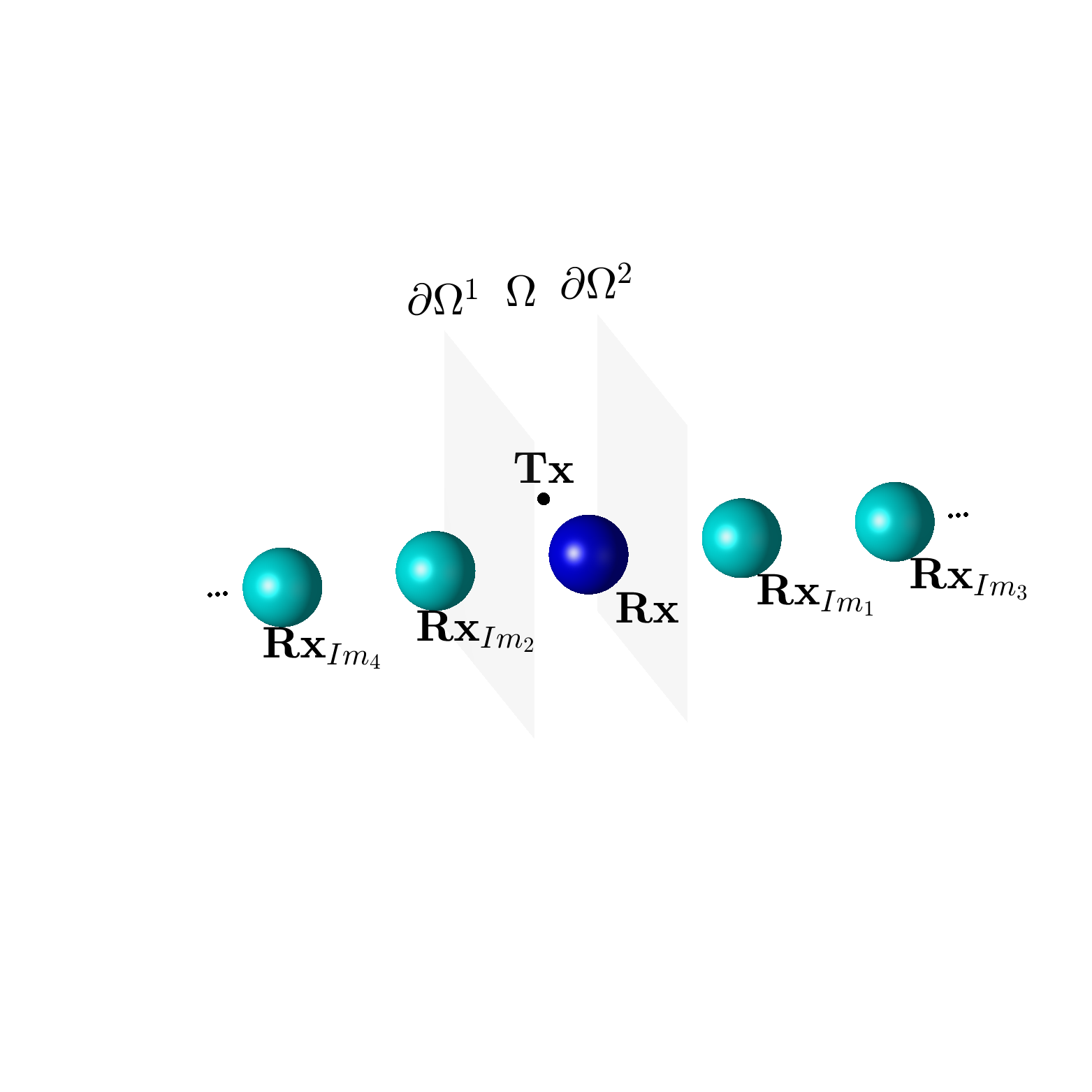}
\caption{\label{fig: Multi Half-Space} Approximated SIMO system when the two infinite reflecting surfaces are omitted.}
\end{figure} 

If we get back to our topology, since Rx and $\text{Rx$_{\text{Im}}$}$ are identical and symmetrically located with respect to the surface, it turns out that
\begin{equation}
    {r_i} = {r_j} = r_r,
\end{equation}
and ${r_0}_{i|j}$ is denoted as ${r_0}_{\text{Rx}|\text{Rx$_{\text{Im}}$}}$, and ${r_0}_{j|i}$ as ${r_0}_{\text{Rx$_{\text{Im}}$}|\text{Rx}}$. By using (18), $p_{\text{hit}}^{\text{SIMO}}(\text{Rx}, t \mid \text{Rx$_{\text{Im}}$})$ can be obtained as

\begin{align}
\begin{split}
    p_{\text{hit}}^{\text{SIMO}}(\text{Rx}, t \mid \text{Rx$_{\text{Im}}$}) &\!=\!  \frac{r_r}{{r_0}} \, \text{erfc}\left( \frac{{r_0} - r_r} {\sqrt{4Dt}} \right)  \
    - \frac{r^{2}_r}{\text{$r_{\text{Im}}$} {r_0}_{\text{Rx}|\text{Rx$_{\text{Im}}$}}} \,  \\ 
    &\!\times\! \text{erfc}\left( -\frac{({r_0} + {r_0}_{\text{Rx}|\text{Rx$_{\text{Im}}$}}) - 2r_r}{\sqrt{4Dt}} \right),
\end{split}
\end{align}
where $p_{\text{hit}}^{\text{SIMO}}(\text{Rx}, t \mid \text{Rx$_{\text{Im}}$},r_0, r_r)$  is denoted as $p_{\text{hit}}^{\text{SIMO}}(\text{Rx}, t \mid \text{Rx$_{\text{Im}}$})$ for simplicity. Finally, by using (16), $p_{\text{hit}}^{\text{SISO}}(\text{Rx}, t \mid r_0, r_r, \partial \Omega)$ can be obtained in closed-form as
\begin{align}
&
    p_{\text{hit}}^{\text{SISO}}(\text{Rx}, t \mid r_0, r_r, \partial \Omega) = \frac{r_r}{{r_0}} \, \text{erfc}\left( \frac{{r_0} - r_r} {\sqrt{4Dt}} \right) \nonumber \\
    &- \frac{r^{2}_r}{\text{$r_{\text{Im}}$} {r_0}_{\text{Rx}|\text{Rx$_{\text{Im}}$}}} \, \times \text{erfc}\left( -\frac{({r_0} + {r_0}_{\text{Rx}|\text{Rx$_{\text{Im}}$}}) - 2r_r}{\sqrt{4Dt}} \right) \nonumber \\
    &+ \frac{r_r}{\text{$r_{\text{Im}}$}} \, \text{erfc}\left( \frac{{\text{$r_{\text{Im}}$}} - r_r} {\sqrt{4Dt}} \right) \nonumber \\
    &- \frac{r^{2}_r}{{r_0} {r_0}_{\text{Rx$_{\text{Im}}$}|\text{Rx}}} \times \text{erfc}\left( -\frac{({\text{$r_{\text{Im}}$}} + {r_0}_{\text{Rx$_{\text{Im}}$}|\text{Rx}}) - 2r_r}{\sqrt{4Dt}} \right).
\end{align}

\subsection{Two Parallel Infinite Reflecting Surfaces}
The previous subsection demonstrates how one can derive the channel response of a SISO system in a half-space with an infinite reflecting surface by utilizing derivations previously proposed for SIMO systems \cite{Gokberk}. Furthermore, we can envision scenarios in which nanonetworks operate within tight regions where the reflective effects from multiple surfaces must be compensated.

From this point, a different topology as shown in Fig. \ref{fig:TwoSurfaces} is considered. A 3-D space is interrupted by two parallel reflecting surfaces that are denoted by $\partial \Omega ^{1}$ and $\partial \Omega ^{2}$. It proceeds with usual notation, where $r_r$ denotes the radius of Rx, and $r_0$ denotes the center-to-center distance between Rx and Tx. For simplicity, Rx is located such that 
\begin{equation}
    d_{min}(\partial \Omega ^{1},\text{Rx}) = d_{min}(\partial \Omega ^{2},\text{Rx}) = d.
\end{equation}

From the Corollary 1, the two parallel infinite reflecting surfaces behave as if they were two parallel mirrors. For every image created by $\partial \Omega ^{1}$, a corresponding image would be created behind $\partial \Omega ^{2}$. Therefore, the number of imaginary receivers is infinite as shown in Fig. \ref{fig: Multi Half-Space}.

In previous work \cite{Gokberk}, the absorption rate of the molecules for each receiver in a multi-receiver system is derived by considering the stealing effect that each receiver imposes on every other receiver. We can define our case as a linear system in Laplace domain as also proposed in \cite{Gokberk}:

\begin{align}
\begin{bmatrix}
1 & -P_{\text{Rx}|{\text{Rx$_{\text{Im$_1$}}$}}}^{\text{SISO}}(s) & \cdots & -P_{\text{Rx}|{\text{Rx$_{\text{Im$_K$}}$}}}^{\text{SISO}}(s) \\
-P_{{\text{Rx$_{\text{Im$_1$}}$}}|\text{Rx}}^{\text{SISO}}(s) & 1 & \cdots & -P_{{\text{Rx$_{\text{Im$_1$}}$}}|{\text{Rx$_{\text{Im$_K$}}$}}}^{\text{SISO}}(s) \\
\vdots & \vdots & \ddots & \vdots \\
-P_{\text{Rx$_{\text{Im$_K$}}$}|\text{Rx}}^{\text{SISO}}(s) & -P_{\text{Rx$_{\text{Im$_K$}}$}|\text{Rx$_{\text{Im$_1$}}$}}^{\text{SISO}}(s) & \cdots & 1 \nonumber
\end{bmatrix}
\end{align}

\begin{align}
\times 
\begin{bmatrix}
P_{\text{Rx}}^{\text{SISO}}(s) \\
P_{\text{Rx$_{\text{Im$_1$}}$}}^{\text{SISO}}(s) \\
\vdots \\
P_{\text{Rx$_{\text{Im$_K$}}$}}^{\text{SISO}}(s) 
\end{bmatrix}
=
\begin{bmatrix}
P_{\text{Rx}}^{\text{SIMO}}(s) \\
P_{\text{Rx$_{\text{Im$_1$}}$}}^{\text{SIMO}}(s) \\
\vdots \\
P_{\text{Rx$_{\text{Im$_K$}}$}}^{\text{SIMO}}(s)
\end{bmatrix},
\end{align}
where $K$ is the number of imaginary receivers. For simplification in notation, assume that Rx corresponds to $\text{Rx$_{\text{Im$_{0}$}}$}$ and center-to-center distance between Rx and Tx, $r_0$, is simply denoted as \text{$r_{\text{Im$_{0}$}}$}. When the inverse Laplace transform of the linear system given in (25) is taken, $p_{\text{hit}}^{\text{SISO}}(\text{Rx}, \text{$r_{\text{Im$_{0}$}}$}, r_r, t \mid \partial \Omega^{1},\partial \Omega^{2})$ can be obtained as

\begin{align}
\begin{split}
    p_{\text{hit}}^{\text{SISO}}&(\text{Rx}, \text{$r_{\text{Im$_{0}$}}$}, r_r, t \mid \partial \Omega^{1},\partial \Omega^{2})   \\ 
    &= \lim_{K \to \infty}    \sum_{i=0}^{K} \frac{r_r}{\text{$r_{\text{Im$_{i}$}}$}} \frac{1}{\sqrt{4\pi Dt}} \frac{(\text{$r_{\text{Im$_{i}$}}$} - r_{r})}{t}  \\
     &\times \exp\left(-\frac{(\text{$r_{\text{Im$_{i}$}}$} - r_r)^2}{4Dt}\right)  \\
   & - \sum_{{j=0, j \ne i}}^{K} \frac{r_r^2}{\text{$r_{\text{Im$_{j}$}}$} {r_0}_{\text{Rx$_{\text{Im$_i$}}$}|\text{Rx$_{\text{Im$_j$}}$}}} \frac{1}{\sqrt{4\pi Dt}}   \\       
   &\times \frac{({\text{$r_{\text{Im$_{j}$}}$}} + {r_0}_{\text{Rx$_{\text{Im$_i$}}$}|\text{Rx$_{\text{Im$_j$}}$}}) - 2r_r}{t}  \\ 
    &\times \exp\left(-\frac{(\text{$r_{\text{Im$_{j}$}}$} + {r_0}_{\text{Rx$_{\text{Im$_i$}}$}|\text{Rx$_{\text{Im$_j$}}$}}) - 4r_r^2}{4Dt} \right).   \\
\end{split}
\end{align}

Unfortunately, (26) cannot be further simplified by elimination of recursively effected molecules as in (18). When $K \rightarrow \infty$, the solution of the linear system gets very complicated.

\section{Performance Evaluation}
For a non-biased evaluation of proposed model, performance evaluation in several different topologies is needed. Every topology parameter must be clear and understandable. For this purpose, the center point of Rx is denoted as a vector, $\Vec{C}$. A point vector $\Vec{\text{Tx}}$ that denotes the position of Tx is defined in 3-D half-space. As stated before, radius of Rx is denoted as $r_r$, minimum distance between $\partial \Omega$ and Rx is denoted as $d$, and center-to-center distance between Rx and Tx is denoted as $r_0$, which is also equal to $d(\Vec{C}$,$\Vec{\text{Tx}})$. To simplify notation, $\partial \Omega$ is located on the $yz$-plane for all topologies. 
\begin{figure}[ht]
  \centering
  \subfloat[Visualization of Topology 0.]{
    \includegraphics[width=0.5\linewidth]{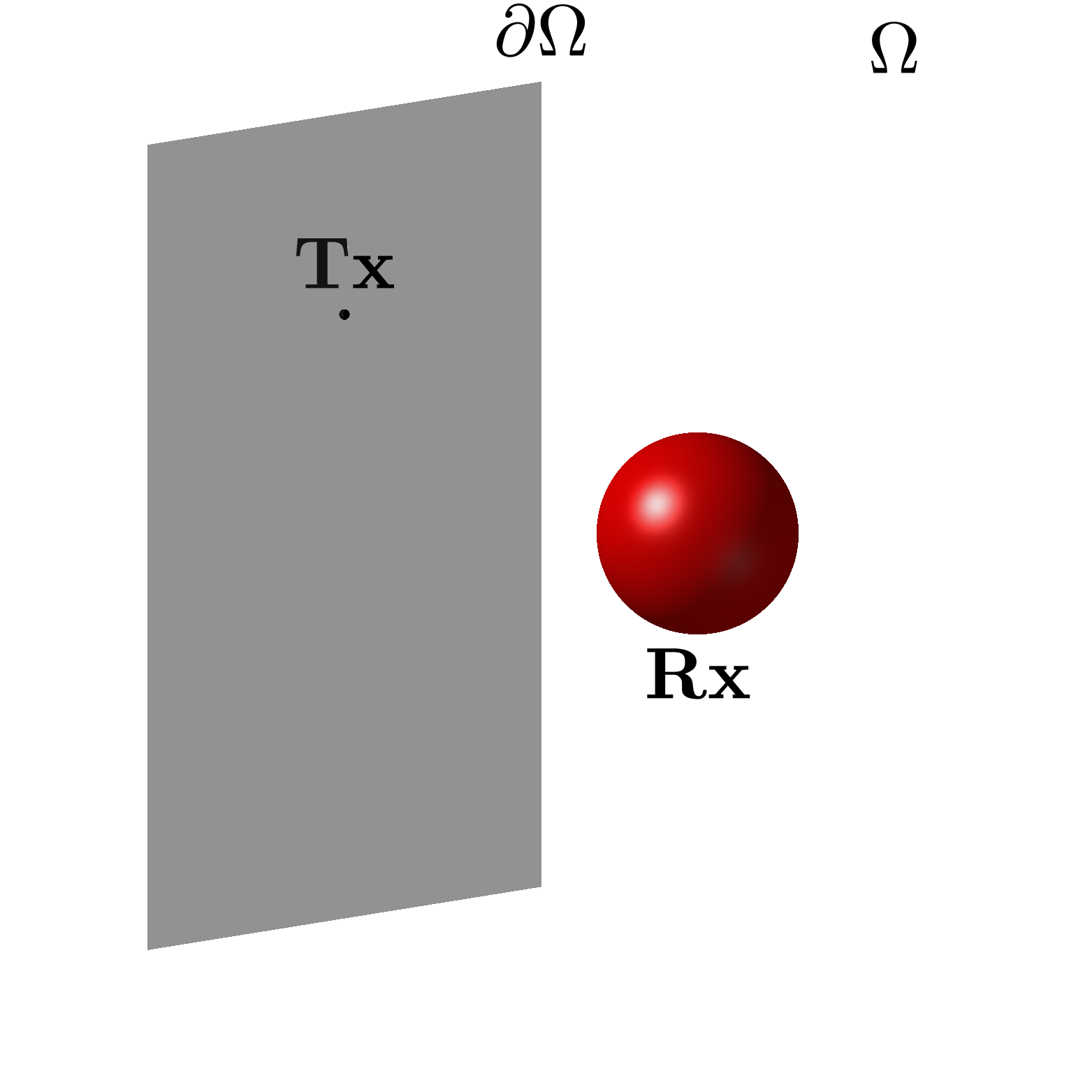}
    \label{fig:top0new}
  } 
  \subfloat[Visualization of Topology 1.]{
    \includegraphics[width=0.5\linewidth]{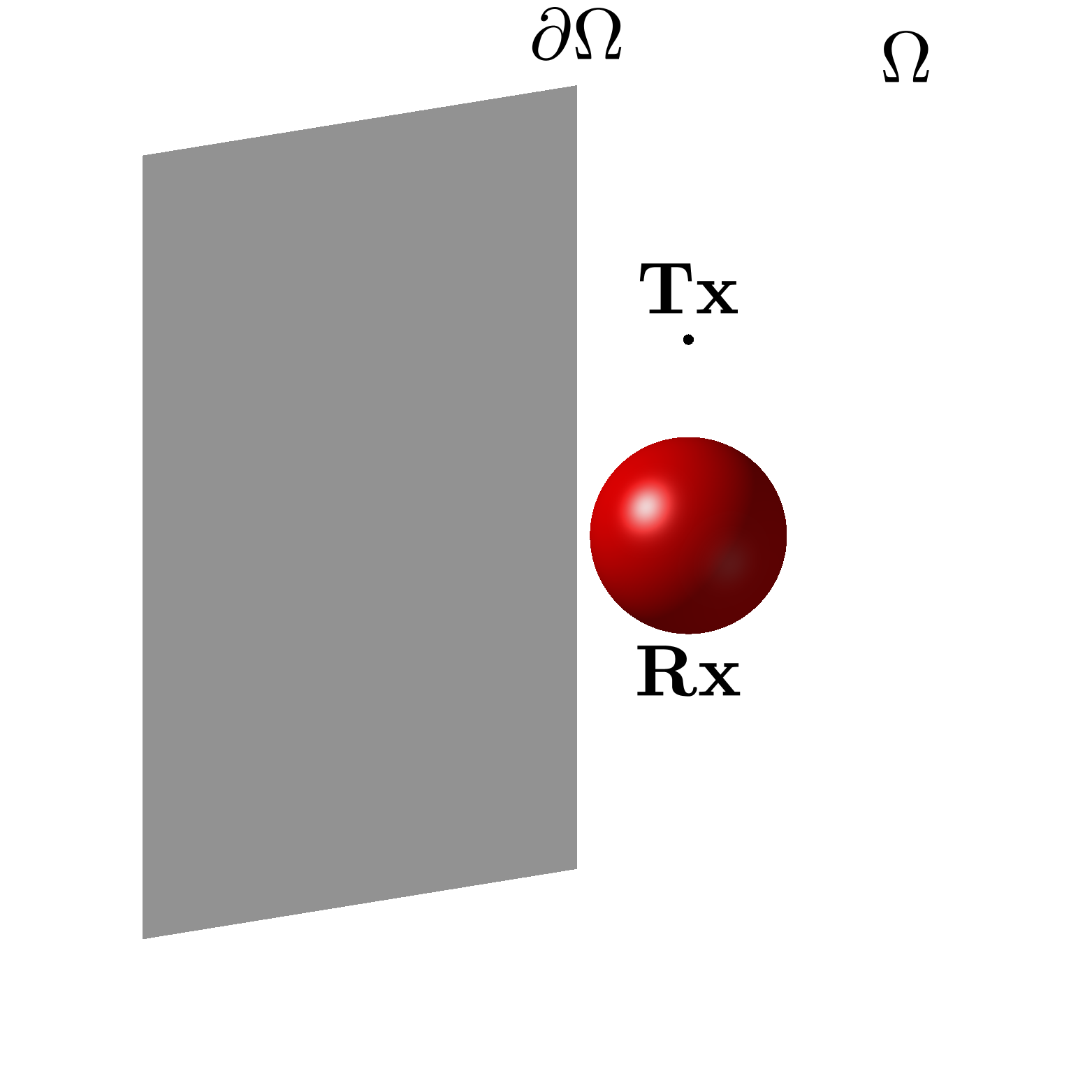}
    \label{fig:top1new}
  }
  \caption{Visualizations of Topology 0 and Topology 1.}
  \label{fig:topologies01}
\end{figure}

\begin{table}[ht]
\renewcommand{\arraystretch}{1.3}
\caption{Simulation Parameters}
\label{Table:Sim_param}
\centering
\begin{tabular}{lcc}
\hline
Parameter & Variable & Value\\
\hline
Diffusion Coefficient & $D$ & $\num{79.4}$  $ {\mu m^{2}/s}$ \\
Simulation Duration & $T$ & ${2}{s}$\\
Time Step & $\Delta t$ & ${10^{-5}}{s}$\\
\hline
\end{tabular}
\end{table}

Channel responses for various topological examples are obtained through a GPU-based simulator that is specialized to simulate different environments and topologies of MCvD systems. Channel response is considered as the hitting rate as defined in \cite{tunabirkan}. The simulations are run 100 times repetitively for each topology with number of molecules of $N = 10^{6}$, and simulation time step size of $\Delta t = 10^{-5}s$. The simulation duration is $T = 2 s.$ At the end of 100 iterations, the channel response is obtained by taking the average to eliminate noise. Performance of the derived analytical model and simulations results are compared in the following subsections.


\subsection{Infinite Reflecting Surface}
\textit{1) Topology 0: Tx in the Vicinity of Reflecting Surface}

To evaluate the performance of the analytical model, one should also consider ``edge-case'' topologies. To see the effect of reflection, when the input is given to the SISO system in the vicinity of reflecting wall, Topology 0 is constructed as shown in Fig. \ref{fig:top0new}. The topology parameters are given in Table \ref{Table0}.
\begin{table}[ht]
\renewcommand{\arraystretch}{1.26}
\caption{\text{Parameters for Topology  0}}
\label{Table0}
\centering
\begin{tabular}{lcc}
\hline
Parameter & Variable & Value\\
\hline
Location of Tx & $\Vec{\text{Tx}}$ &  $ (0,0,10 \mu m)$ \\
Location of Rx & $\Vec{C}$ & $(10 \mu m,0,0)$\\
Radius of Rx & $r_r$ & $3 \mu m, 5\mu m, 8\mu m$\\
\hline
\end{tabular}
\end{table}

\begin{figure*}[ht]
    \centering
  \begin{minipage}{0.25\linewidth}
  \centering
   \subfloat[Plot of cumulative distribution of \\ $p_{\text{hit}}^{\text{SISO}}$ for Topology 0.\label{Top0CDF}]{\includegraphics[width=1\columnwidth]{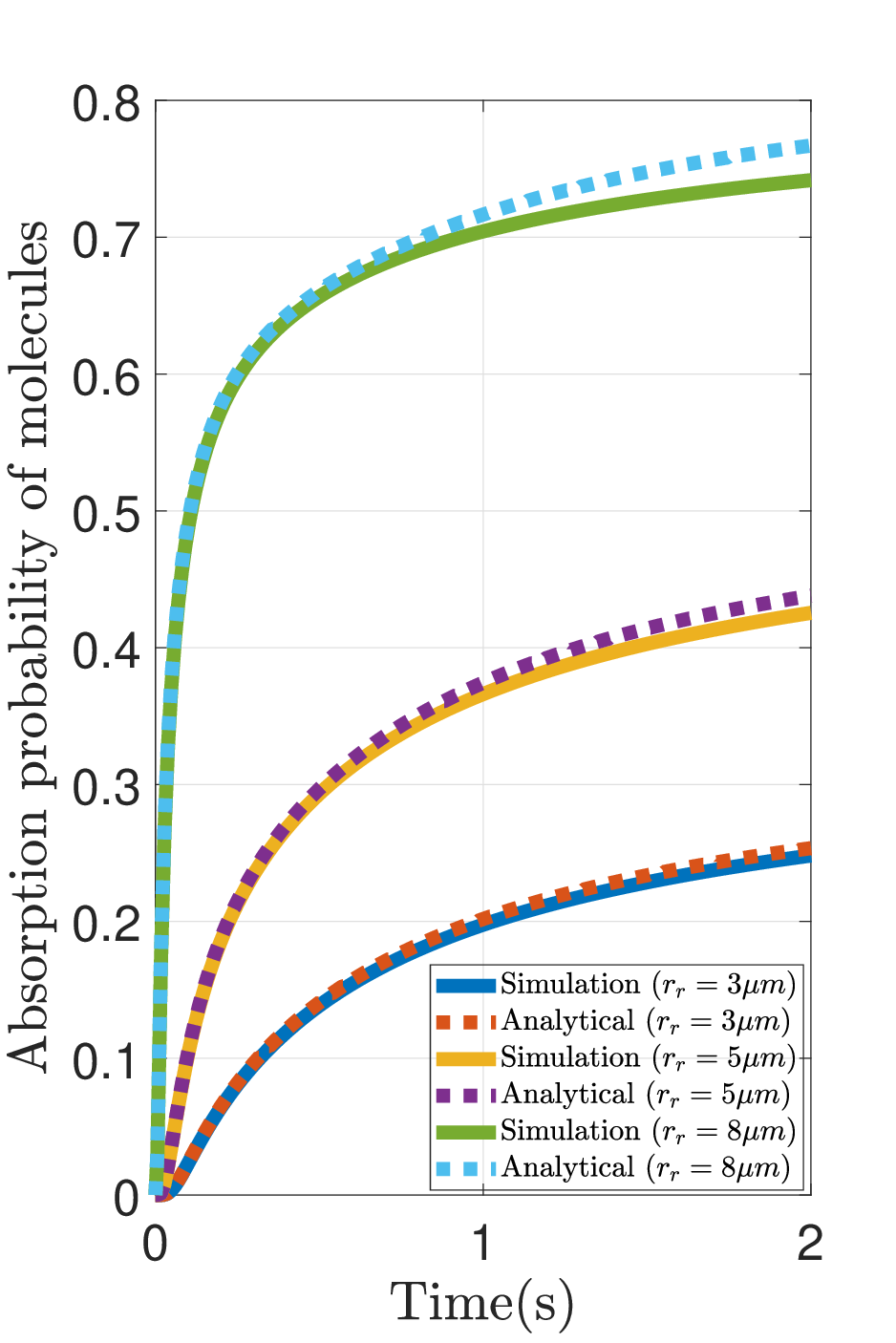}}
  \end{minipage}%
  \begin{minipage}{0.25\linewidth}
  \centering
   \subfloat[Plot of $p_{\text{hit}}^{\text{SISO}}$ for Topology 0 \\ ($r_r = 5\mu m$).\label{Top0PDF}]{\includegraphics[width=1\columnwidth]{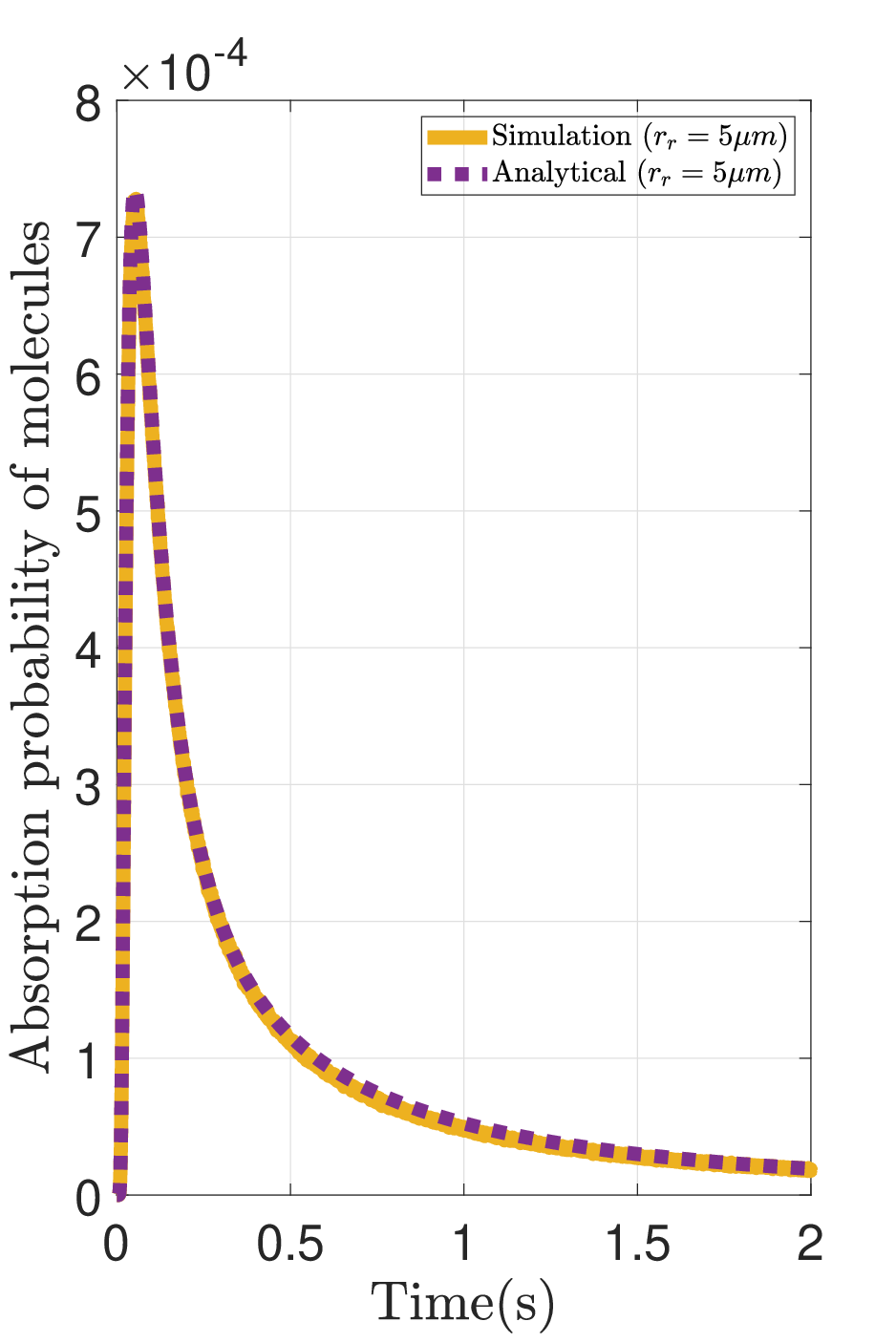}}
  \end{minipage}%
  \begin{minipage}{0.25\linewidth}
  \centering
  \subfloat[Plot of cumulative distribution of \\ $p_{\text{hit}}^{\text{SISO}}$ for Topology 1.\label{Top1CDF}]{\includegraphics[width=1\columnwidth]{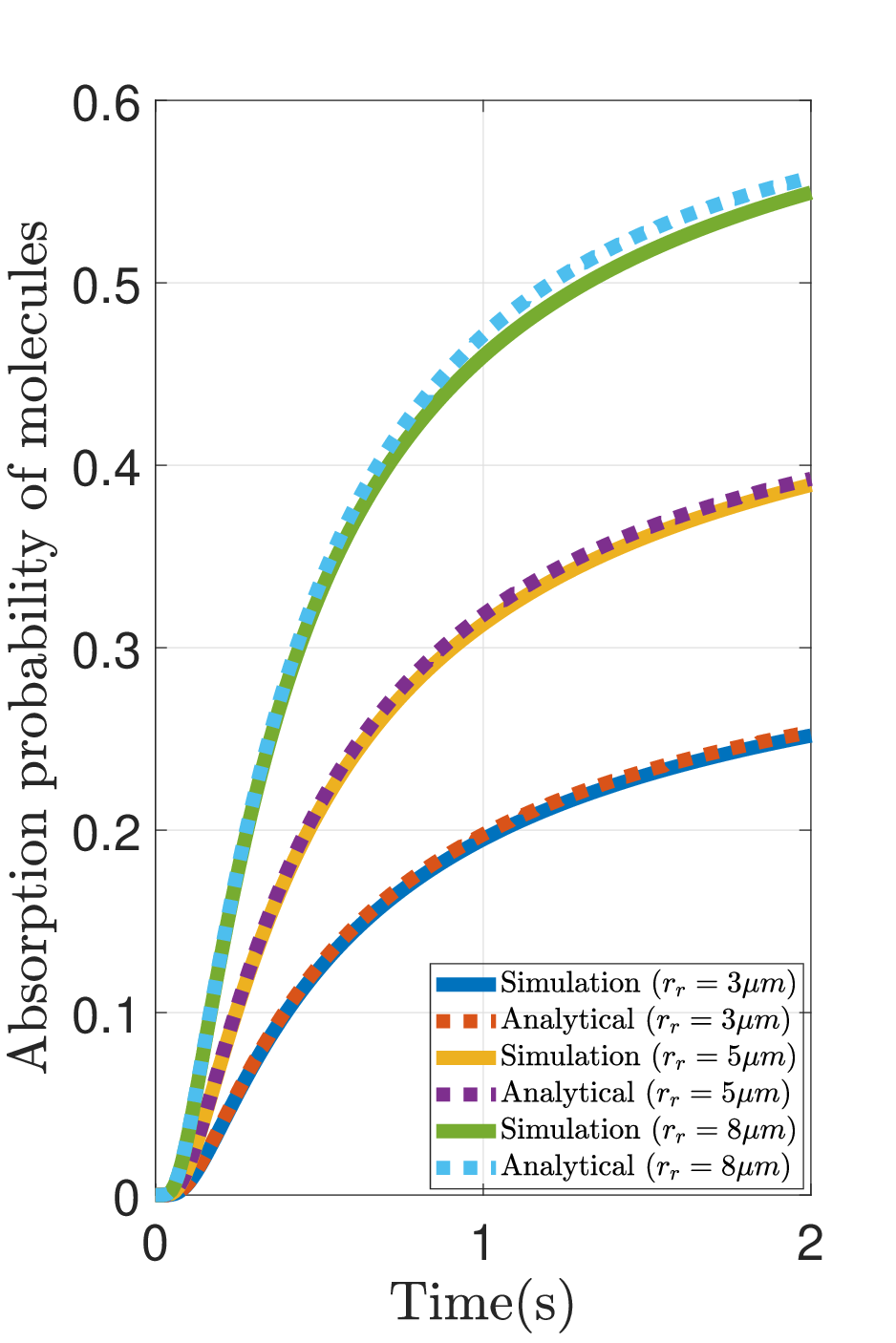}}
  
  \end{minipage}%
  \begin{minipage}{0.25\linewidth}
  \centering
  \subfloat[Plot of $p_{\text{hit}}^{\text{SISO}}$ for Topology 1 \\ ($r_r = 5\mu m$).\label{Top1Pdf}]{\includegraphics[width=1\columnwidth]{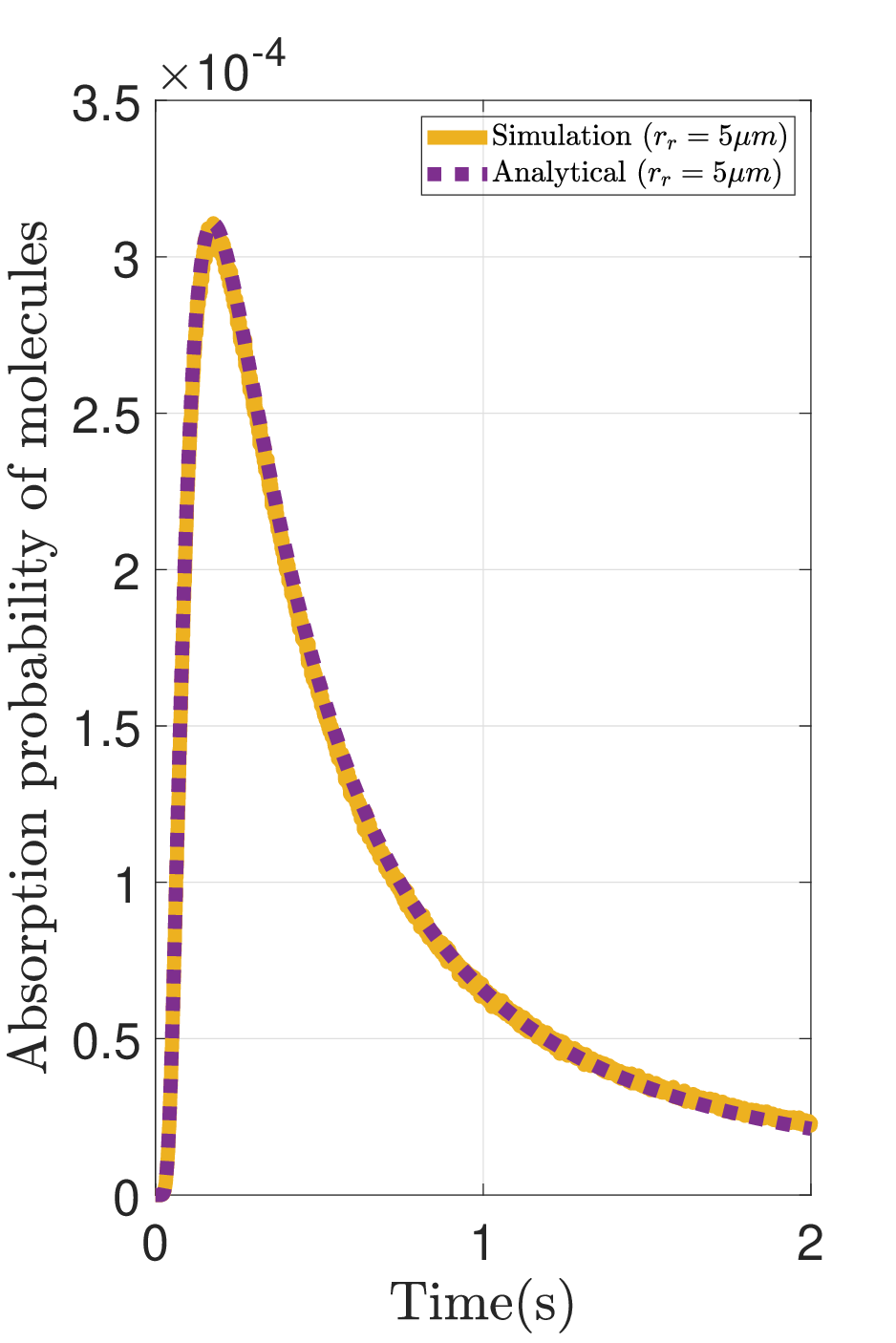}}
  
  \end{minipage}%
  \caption{Plots of absorption probabilities of molecules by Rx with radii $3 \mu m, 5 \mu m, $ and $ 8 \mu m $ for Topology 0 and Topology 1. \label{All_results} }
\end{figure*}



\begin{table}[ht]
\renewcommand{\arraystretch}{1.1}
\caption{RMSE for Topology 0}
\label{table0rms}
\centering
\begin{tabular}{c||c||c}
\hline
\bfseries $r_{r} = 3 \mu m$ & \bfseries $r_{r} = 5 \mu m$ & \bfseries $r_{r} = 8 \mu m$ \\
\hline\hline
0.0040 & 0.0085 & 0.0150 \\
\hline
\end{tabular}
\end{table}

Performance of Topology  0 is evaluated by varying the radius of Rx as $3 \mu m, 5\mu m, $ and $ 8\mu m$, respectively. All information required to analyze Topology  0 is given in Table \ref{Table0}, from which $d$ and $r_0$ can be derived. For instance, for $r_r = 8 \mu m$, $d$ equals  $2 \mu m$ while for $r_r = 5 \mu m$, $d$ equals $5 \mu m$. Additionally, $r_0$ is $10\sqrt{2} \mu m$ for all radii. As a result, the minimum distance between Rx and Tx, $r_0 - r_r$ changes with varying radii. The RMSE performance of the proposed model with different Rx radii is given in Table \ref{table0rms}.

\textit{2) Topology 1: Side-Mirror Effect}

Topology  1 is constructed to evaluate the performance of the analytical solution under the influence of the side mirror effect as shown in Fig. \ref{fig:top1new}. All information required to analyze Topology  1 is given in Table \ref{Top1table}, from which $d$ and $r_0$ can be derived. For example, for $r_r = 8 \mu m$, $d$ equals  $2 \mu m$ while for $r_r = 5 \mu m$, $d$ equals $5 \mu m$. Additionally, $r_0$ is $10 \mu m$ for all radii. As a result, the minimum distance between Rx and Tx, $r_0 - r_r$ changes with varying radii. Performance of Topology  1 is evaluated by varying the radius of Rx as $3 \mu m, 5\mu m,$ and $ 8\mu m$, respectively. The RMSE performance of the proposed model with different Rx radii is given in Table \ref{Top1RMS}.

\begin{table}[ht]
\renewcommand{\arraystretch}{1.25}
\caption{\text{Parameters for Topology  1}}
\label{Top1table}
\centering
\begin{tabular}{lcc}
\hline
Parameter & Variable & Value\\
\hline
Location of Tx & $\Vec{\text{Tx}}$ &  $ (10 \mu m,0,10 \mu m)$ \\
Location of Rx & $\Vec{C}$ & $(10 \mu m,0,0)$\\
Radius of Rx & $r_r$ & $3 \mu m, 5\mu m, 8\mu m$\\
\hline
\end{tabular}
\end{table}

\begin{table}[ht]
\renewcommand{\arraystretch}{1.1}
\caption{RMSE for Topology 1}
\label{Top1RMS}
\centering
\begin{tabular}{c||c||c}
\hline
\bfseries $r_{r} = 3 \mu m$ & \bfseries $r_{r} = 5 \mu m$ & \bfseries $r_{r} = 8 \mu m$ \\
\hline\hline
0.0021 & 0.0040 & 0.0091 \\
\hline
\end{tabular}
\end{table}

\begin{figure}[ht]
  \centering
  \subfloat[Visualization of Topology 2.]{
    \includegraphics[width=0.5\linewidth]{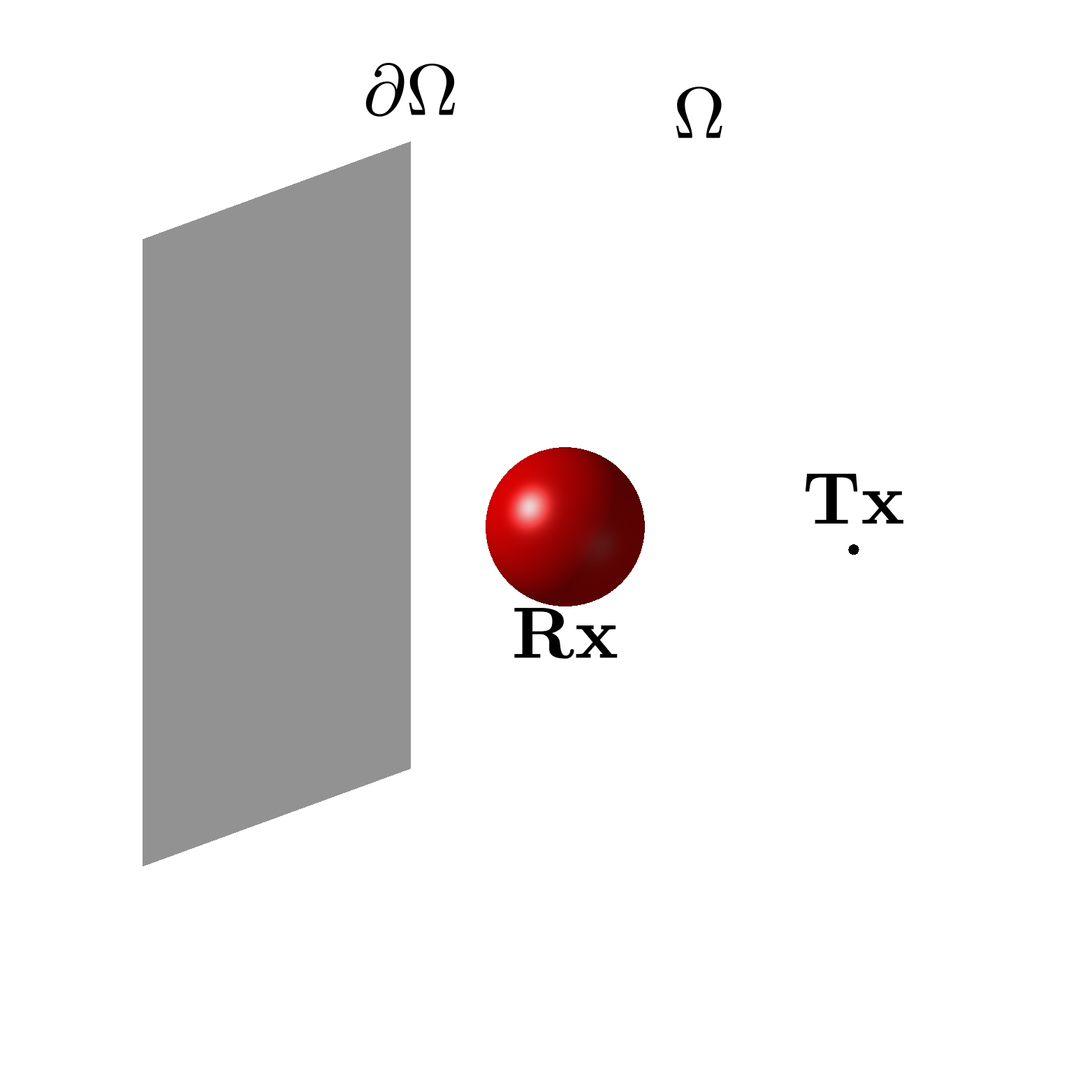}
    \label{fig:top0}
  } 
  \subfloat[Visualization of Topology 3.]{
    \includegraphics[width=0.5\linewidth]{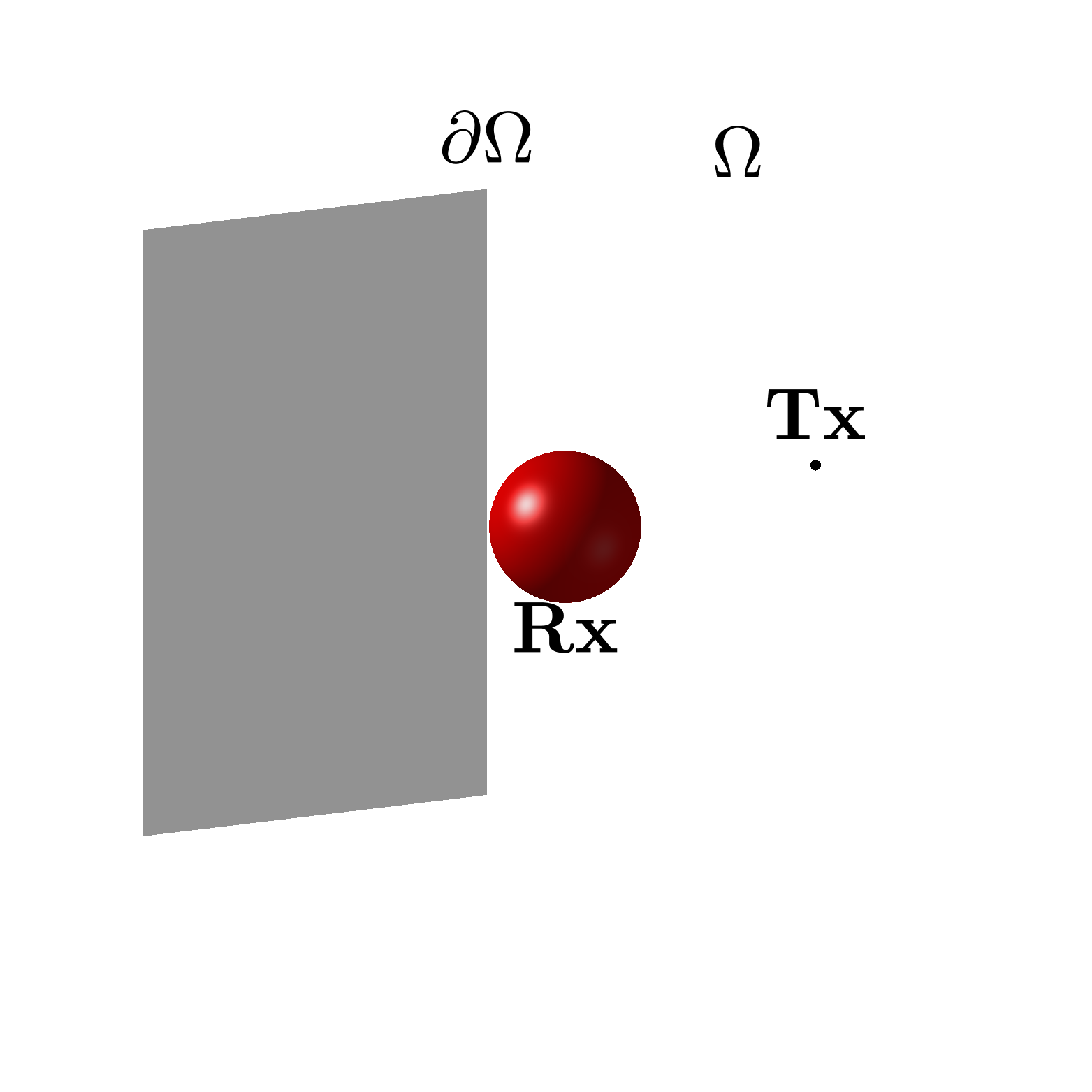}
    \label{fig:top1}
  }
   \caption{Visualizations of Topology 2 and Topology 3.}
  \label{fig:topologies23}
\end{figure}

\textit{3) Topology  2: Total Eclipse Effect}

Derivation of the analytical solution to the channel response of MCvD systems in 3-D half-space with an infinite reflecting surface relies on the channel modeling for multi-receiver systems. However, due to the recursive nature of channel modeling, these systems are erroneous when one of the receiver is totally eclipsed by an other receiver \cite{Gokberk}. Since our proposed model sums all the absorption rates of molecules to each of Rxs, it ignores the stealing effect. As a result, the idea presented in (16) gives robustness to our model in total eclipse. Moreover, Topology 2, as shown in Fig. \ref{fig:top0}, could be used to see exactly the effect of $d$ on our proposed model. Therefore, the performance of Topology  2 is evaluated with different $d$ values of $1 \mu m, 3\mu m,$ and $ 5\mu m$, respectively. In order to increase or decrease the parameter $d$, $\partial \Omega$ is shifted on the $x$-axis accordingly while $r_0$ and $r_r$ remain constant. Topology parameters are given in Table \ref{Top2T}. The RMSE performance of the proposed model with different $d$ values is given in Table \ref{top2rms}.

\begin{table}[ht]
\renewcommand{\arraystretch}{1.25}
\caption{\text{Parameters for Topology  2}}
\label{Top2T}
\centering
\begin{tabular}{lcc}
\hline
Parameter & Variable & Value\\
\hline
Location of Tx & $\Vec{\text{Tx}}$ &  $ (20 \mu m,0,0)$ \\
Location of Rx & $\Vec{C}$ & $(10\mu m,0,0)$\\
Radius of Rx & $r_r$ & $5\mu m$\\
$d_{min}(\partial \Omega,\text{Rx})$ & $d$ & $1 \mu m, 3\mu m, 5\mu m$\\
\hline
\end{tabular}
\end{table}

\begin{table}[h!]
\renewcommand{\arraystretch}{1.1}
\caption{RMSE for Topology 2}
\label{top2rms}
\centering
\begin{tabular}{c||c||c}
\hline
\bfseries $d = 1 \mu m$ & \bfseries $d = 3 \mu m$ & \bfseries $d = 5 \mu m$ \\
\hline\hline
0.0088 & 0.0039 & 0.0048 \\
\hline
\end{tabular}
\end{table}

\begin{figure}[ht]
  \centering
  \includegraphics[width=0.5\linewidth]{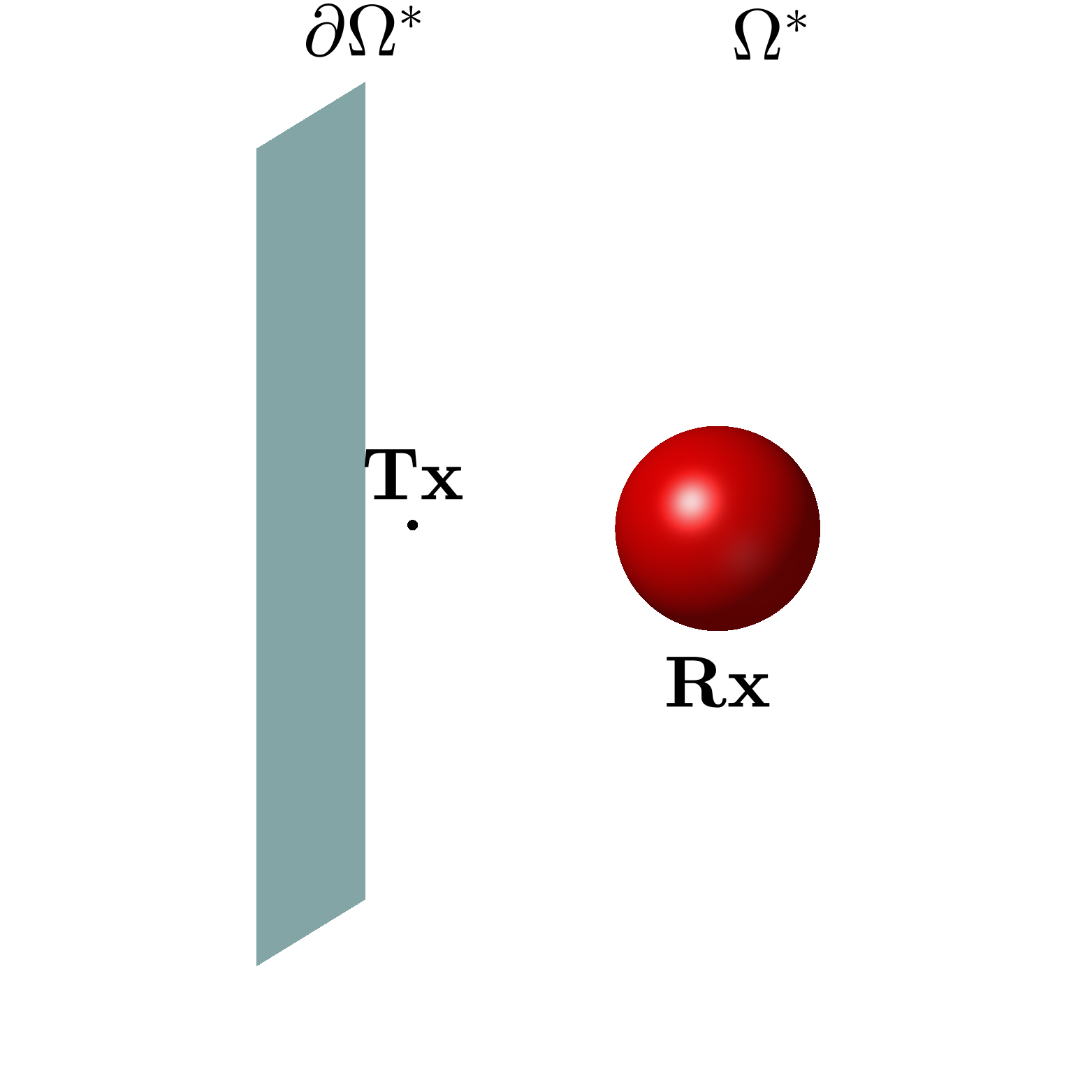}
  \caption{Visualization of Topology 4}
  \label{fig:top4}
    
\end{figure}

\begin{figure*}[ht]
\centering
  \begin{minipage}{0.25\linewidth}
  \centering
  \subfloat[Plot of cumulative distribution of \\ $p_{\text{hit}}^{\text{SISO}}$ for Topology 2.\label{Top2CDF}]{\includegraphics[width=1\columnwidth,keepaspectratio]{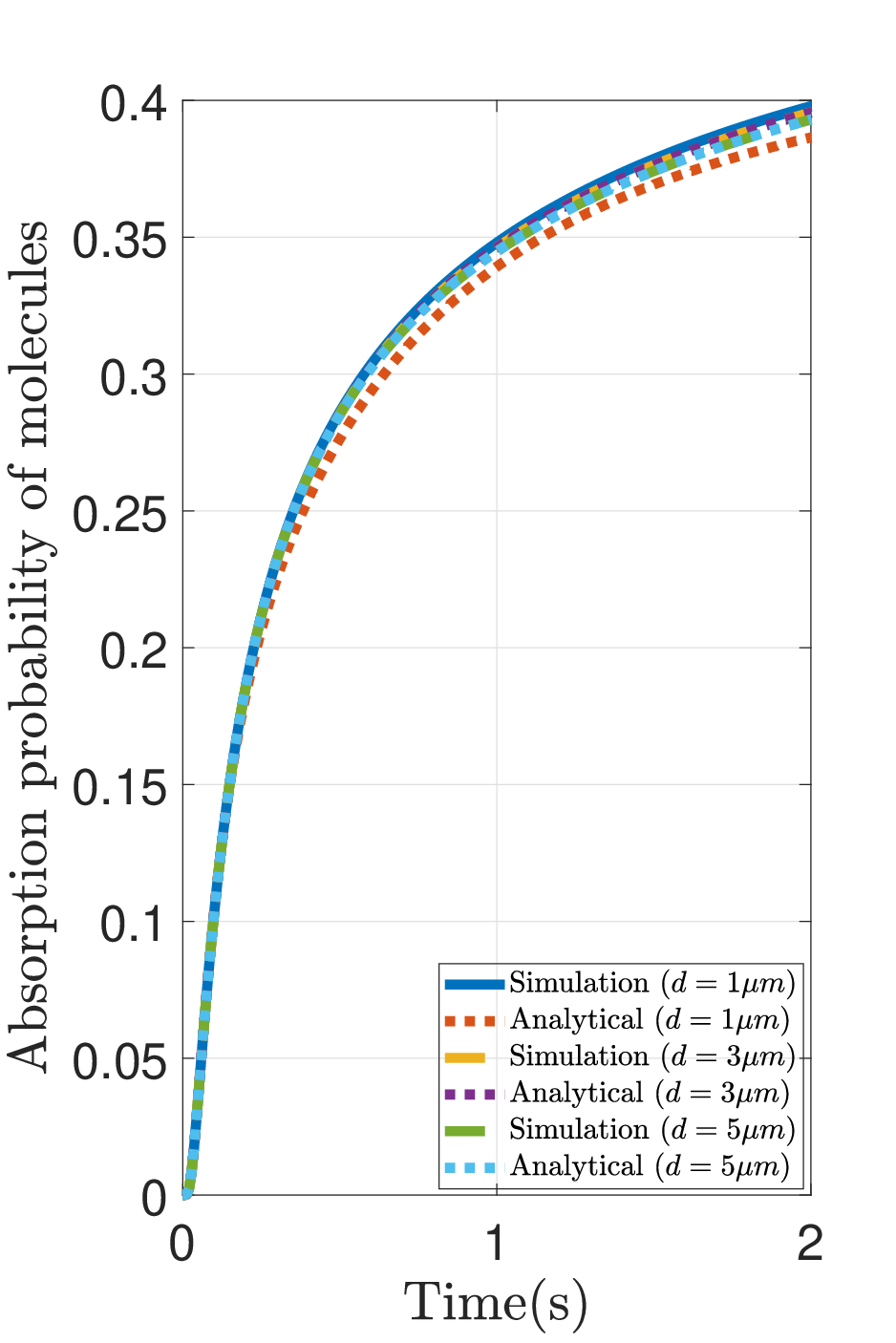}}
  \end{minipage}%
  \begin{minipage}{0.25\linewidth}
  \centering
  \subfloat[Plot of $p_{\text{hit}}^{\text{SISO}}$ for Topology 2 \\ ($r_r = 5\mu m, d = 1 \mu m$).\label{Top2PDF}]{\includegraphics[width=1\columnwidth,keepaspectratio]{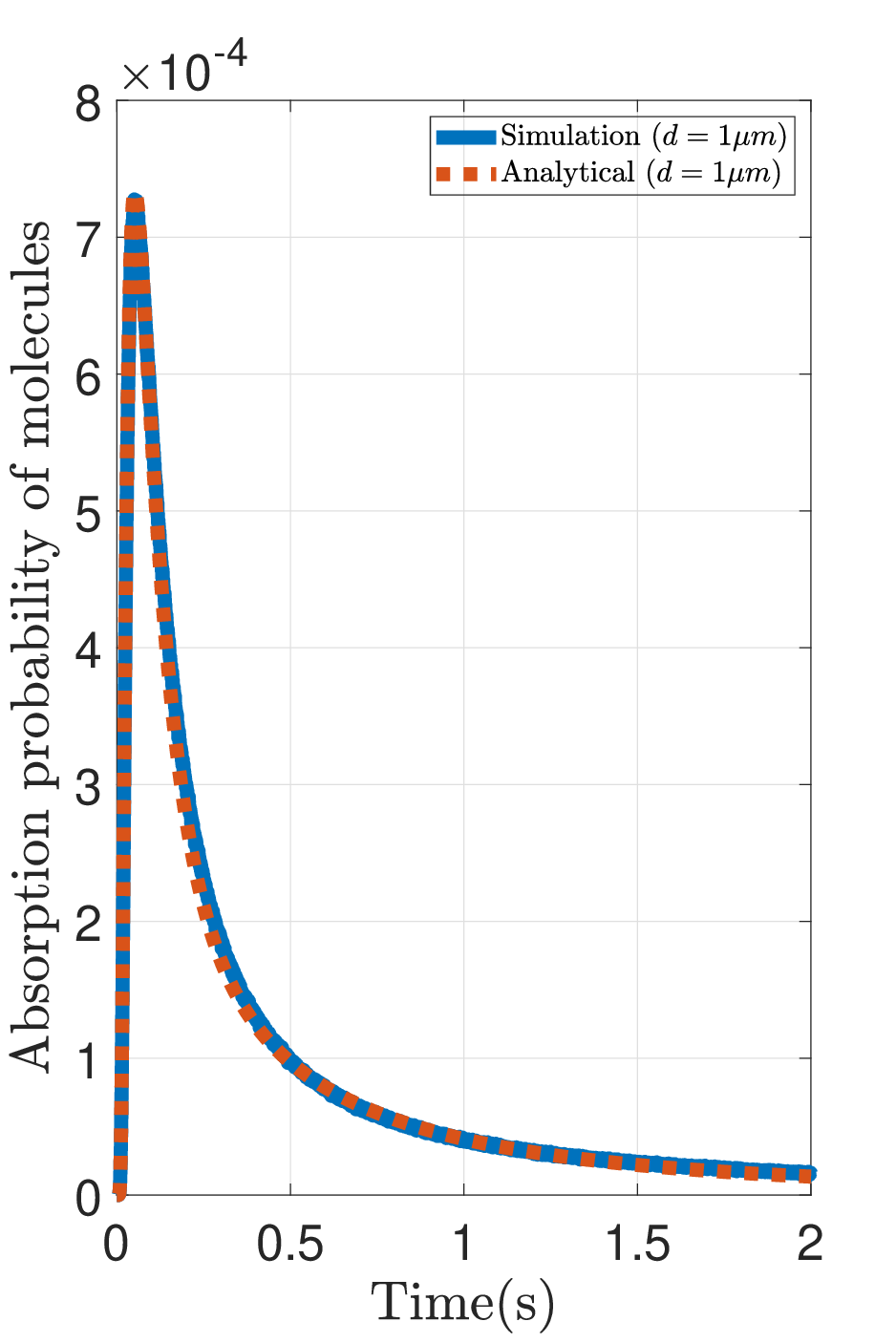}}
  \end{minipage}%
  \begin{minipage}{0.25\linewidth}
  \centering
  \subfloat[Plot of cumulative distribution of \\ $p_{\text{hit}}^{\text{SISO}}$ for Topology 3.\label{Top3CDF}]{\includegraphics[width=1\columnwidth,keepaspectratio]{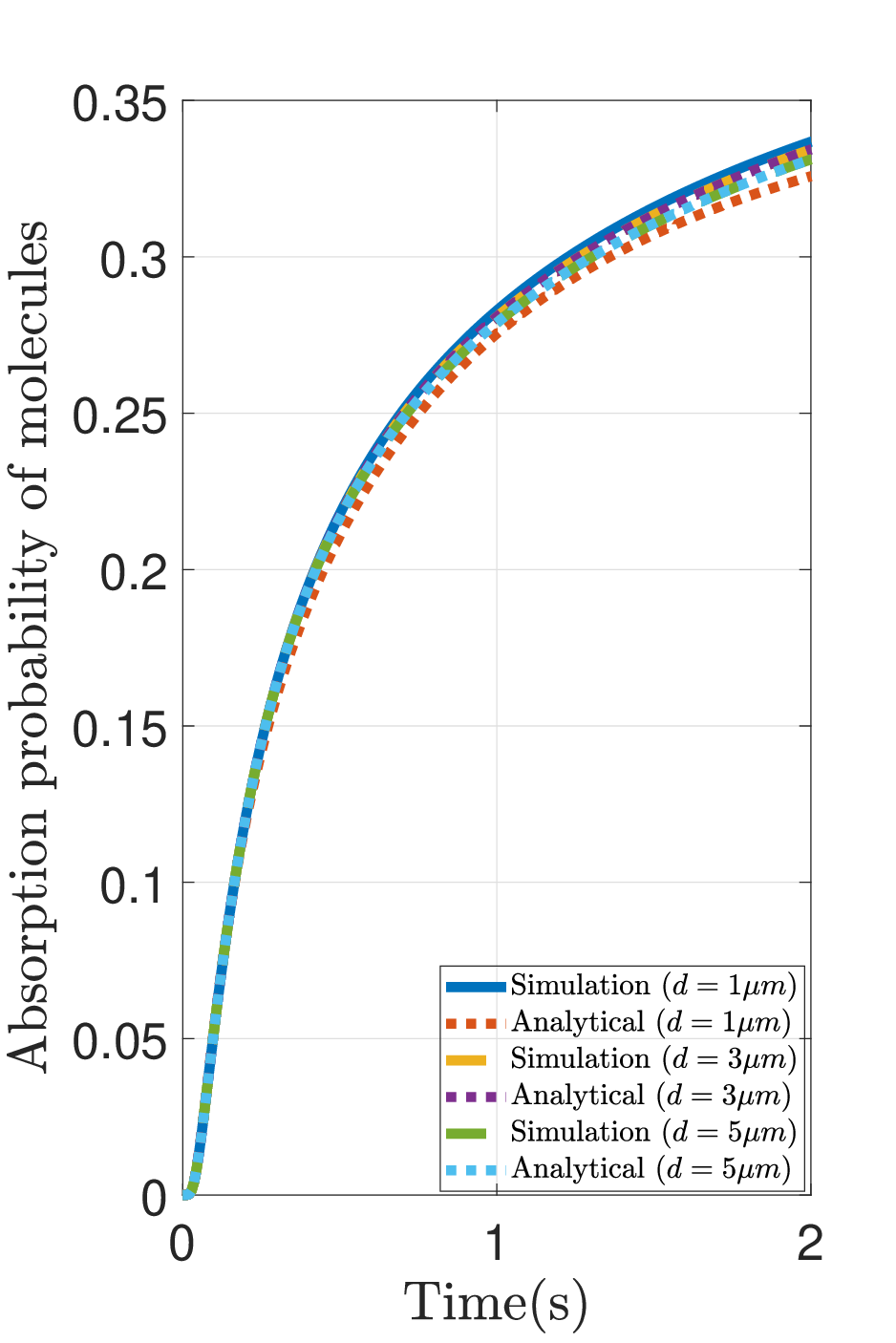}}
  \end{minipage}%
  \begin{minipage}{0.25\linewidth}
  \centering
  \subfloat[Plot of $p_{\text{hit}}^{\text{SISO}}$ for Topology 3 \\ ($r_r = 5\mu m, d = 1 \mu m$).\label{Top3PDF}]{\includegraphics[width=1\columnwidth,keepaspectratio]{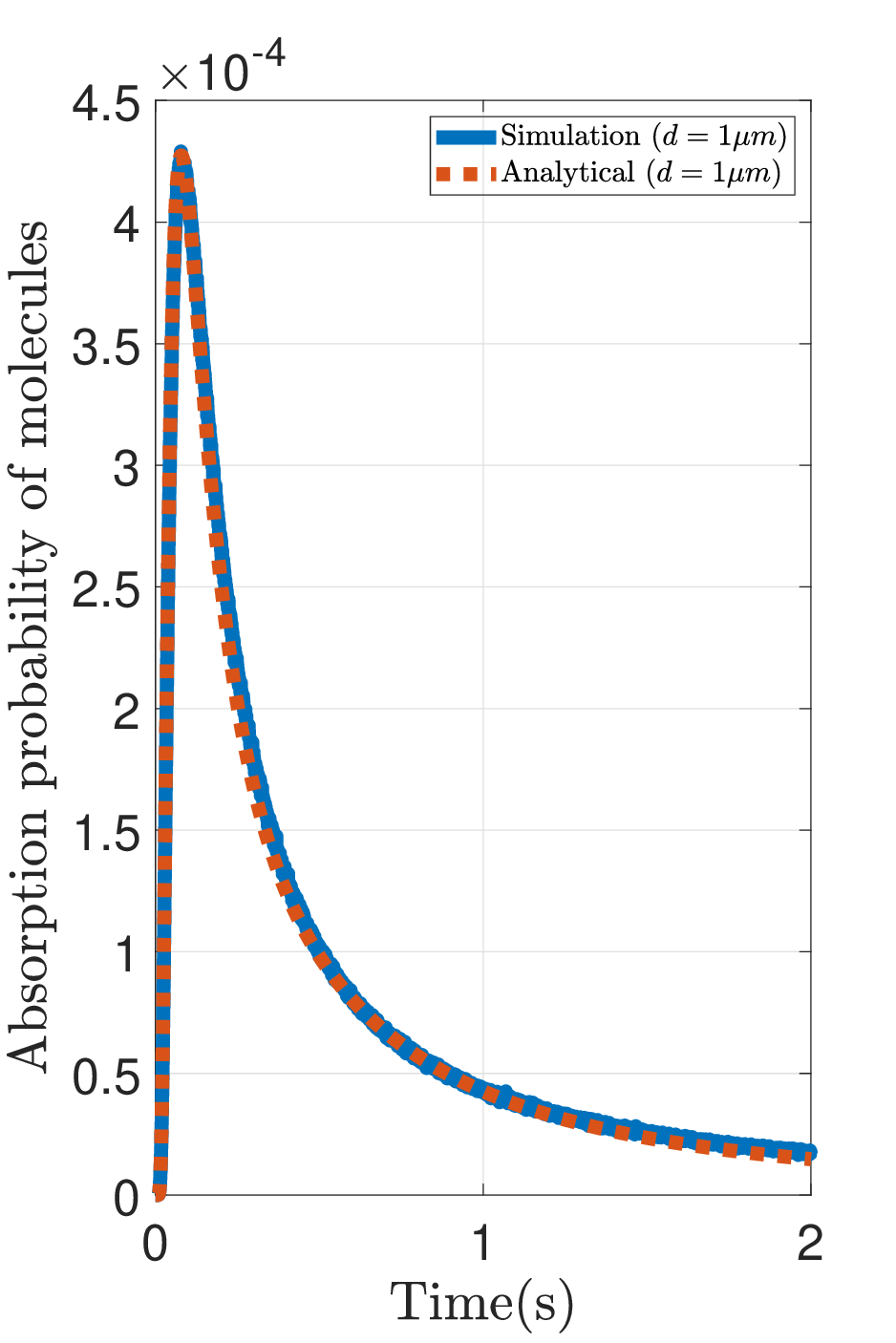}}
  \end{minipage}%
  
  \caption{Plot of absorption probabilities of molecules by Rx, where $d$ equals to $1 \mu m, 3 \mu m, $ and $5 \mu m $, for Topology 2 and Topology 3. \label{Topology23figurelabel} }
\end{figure*}

\textit{4) Topology  3: Half Eclipse Effect}

To see the performance of our model at half-eclipse angle, which is stated in \cite{Gokberk}, Topology  3 is constructed as shown in Fig. \ref{fig:top1}. Tx is shifted by 5 $\mu m$ in the $+z$ direction from its location in Topology 2. Topology parameters are given in Table \ref{Table3}, and performance of Topology  3 is evaluated with different $d$ values that are $1 \mu m, 3\mu m,$ and $5\mu m,$ respectively. In order to increase or decrease the parameter $d$, $\partial \Omega$ is shifted on the $x$-axis accordingly while $r_0$ and $r_r$ remain constant. The RMSE performance of the proposed model with different $d$ values is given in Table \ref{top3rms}.

\subsection{Finite Reflecting Surface}
\begin{figure}[ht]
  \centering
  \subfloat[Plots of cumulative distribution \\ of  $p_{\text{hit}}^{\text{SISO}}(\text{Rx},t \mid r_0,r_r,\partial \Omega)$ for \\ Topology 2  with a finite reflecting \\ surface, and Topology 4.]{
    \includegraphics[width=0.5\linewidth,keepaspectratio]{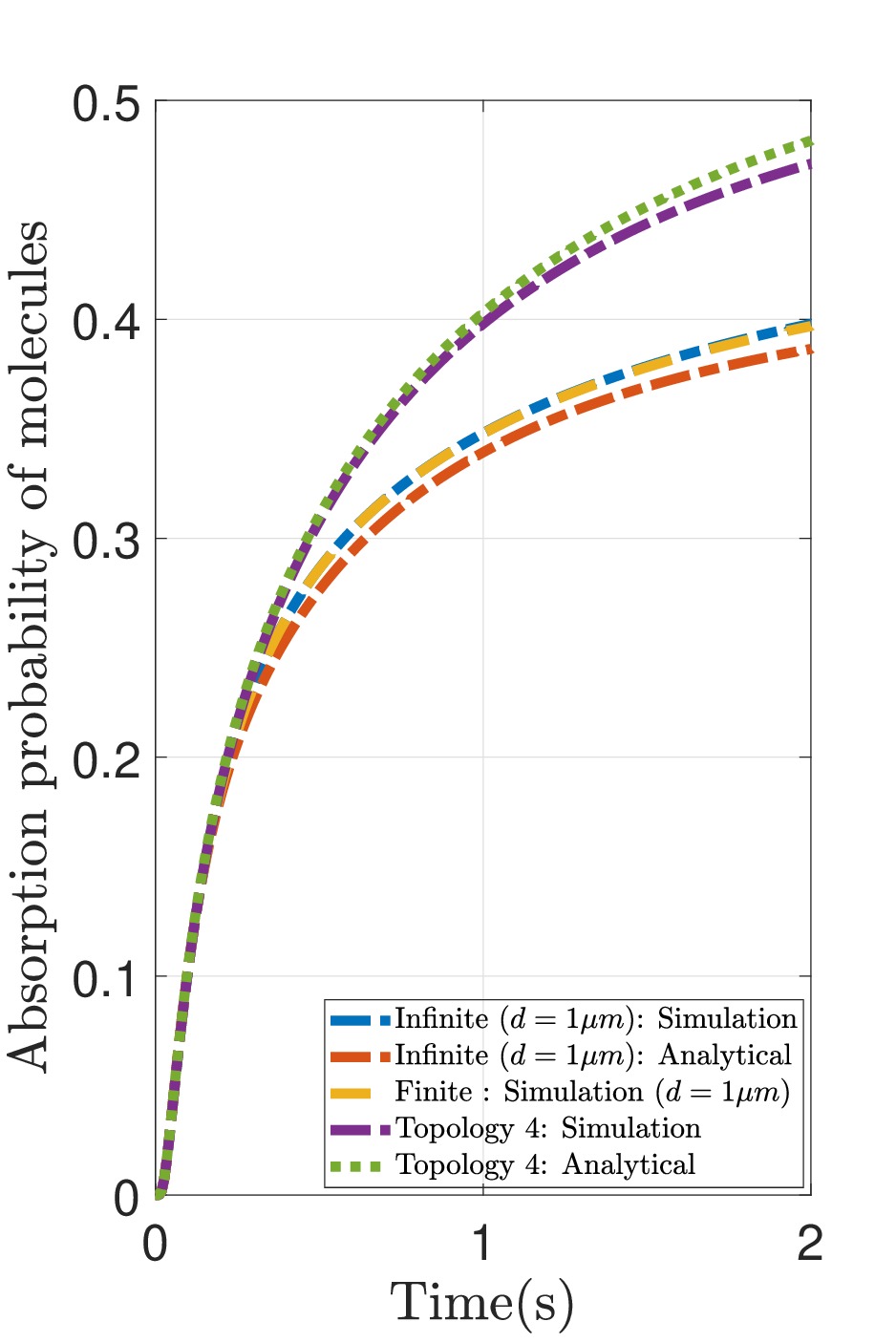}
    \label{fig:finite}
  } 
  \subfloat[Plot of cumulative distribution \\ of  $p_{\text{hit}}^{\text{SISO}} (\text{Rx}, \text{$r_{\text{Im$_{0}$}}$}, r_r, t \mid \partial \Omega^{1},\partial \Omega^{2})$   $ \\(r_{r} = 5 \mu m, d = 3 \mu m$).]{\includegraphics[width=0.5\linewidth,keepaspectratio]{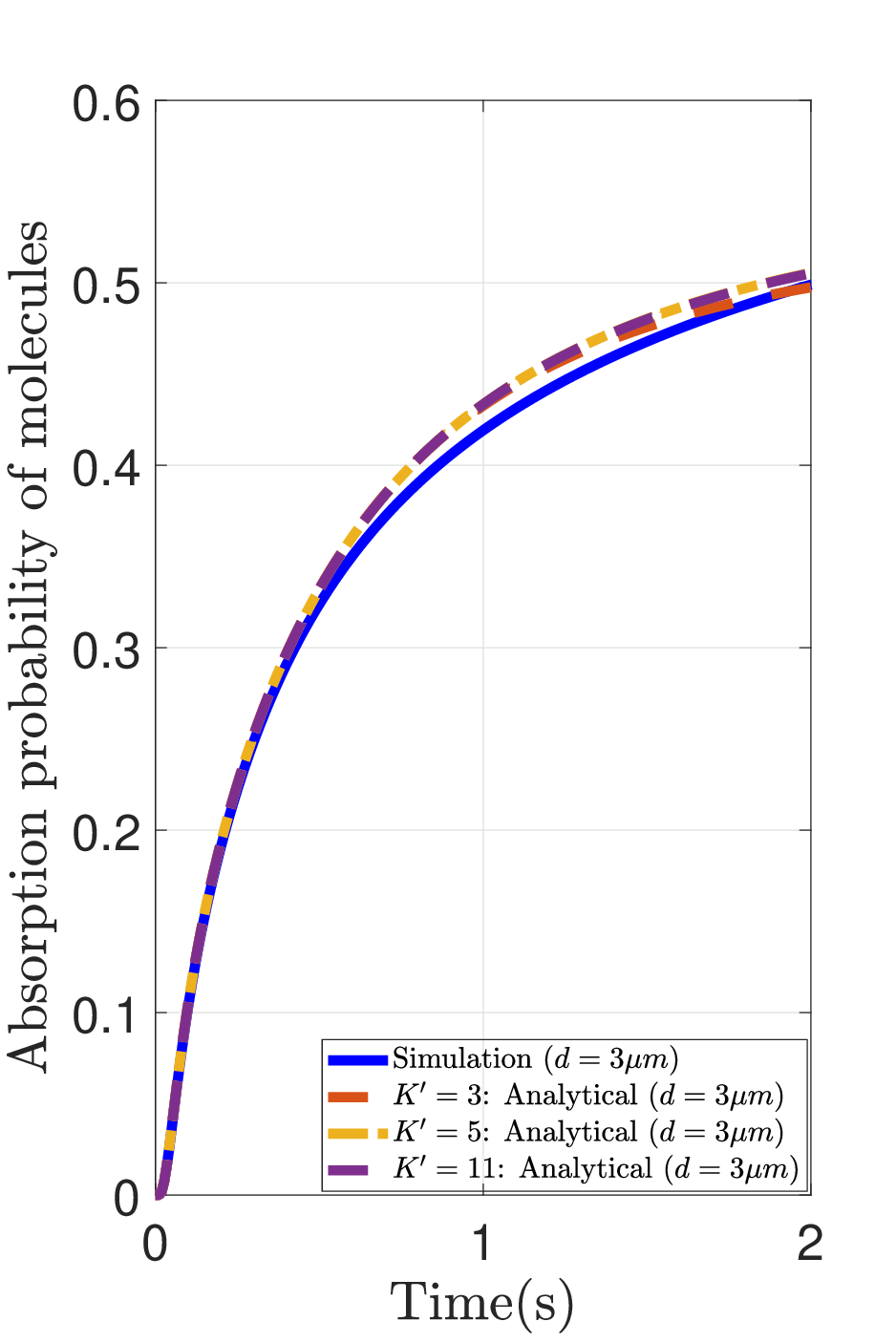}
    \label{fig:double}}
  \caption{(a) Plots of absorption probability of molecules by Rx in 3-D half-space with a finite reflecting surface with respect to time. (b) Performance evaluation
of a MCvD channel bounded by two infinite parallel surfaces as shown in Fig. 3.}
  \label{doubleandfinitefig}
\end{figure}

It is observed that the intuition behind the derivation of (16) holds its prominence when reflecting surface is finite, if it is sufficiently wide compared to the size of Rx. Topology 2 is reconstructed, when $d = 1 \mu m$, with a finite reflecting surface instead of the infinite reflecting surface. The constructed finite reflecting surface is a thin square reflecting layer whose sides are $40 \mu m$. The remaining topology parameters are same as given in Table \ref{Top2T}. The RMSE for Topology 2 with the finite reflecting surface is 0.0085. The performance analysis between an infinite reflecting surface and finite reflecting surface is shown as Fig. \ref{fig:finite}. To see the further effects of finite reflecting surface, Topology 4 is constructed. A square surface, whose sides are $40 \mu m$ and  center of mass is at origin, is located on the $yz$-plane as shown in Fig. \ref{fig:top4}. $\partial \Omega ^{*}$ denotes the finite reflecting surface. Topology parameters are given in Table \ref{Table4}.  The RMSE for Topology 4 is 0.0060.  

\begin{table}[h!]
\renewcommand{\arraystretch}{1.25}
\caption{\text{Parameters for Topology  3}}
\label{Table3}
\centering
\begin{tabular}{lcc}
\hline
Parameter & Variable & Value\\
\hline
Location of Tx & $\Vec{\text{Tx}}$ &  $ (20 \mu m,0,5 \mu m)$ \\
Location of Rx & $\Vec{C}$ & $(10\mu m,0,0)$\\
Radius of Rx & $r_r$ & $5\mu m$\\
$d_{min}(\partial \Omega,\text{Rx})$ & $d$ & $1 \mu m, 3\mu m, 5\mu m$\\
\hline
\end{tabular}
\end{table}

\begin{table}[h!]
\renewcommand{\arraystretch}{1.1}
\caption{RMSE for Topology 3}
\label{top3rms}
\centering
\begin{tabular}{c||c||c}
\hline
\bfseries $d = 1 \mu m$ & \bfseries $d = 3 \mu m$ & \bfseries $d = 5 \mu m$ \\
\hline\hline
0.0076 & 0.0030 & 0.0039 \\
\hline
\end{tabular}
\end{table}

\begin{table}[h!]
\renewcommand{\arraystretch}{1.25}
\caption{\text{Parameters for Topology  4}}
\label{Table4}
\centering
\begin{tabular}{lcc}
\hline
Parameter & Variable & Value\\
\hline
Location of Tx & $\Vec{\text{Tx}}$ &  $ (5 \mu m,0,0 \mu m)$ \\
Location of Rx & $\Vec{C}$ & $(15\mu m,0,0)$\\
Radius of Rx & $r_r$ & $5\mu m$\\
\hline
\end{tabular}
\end{table}

\subsection{Bounded Space with Two Infinite Reflecting Surface}
As shown in Fig. \ref{fig:TwoSurfaces}. when a SISO system in 3-D space is bounded by two infinite reflecting surfaces, the analytical solution is given in equation (26). However, the solution is intractable when $K$ $\rightarrow \infty$. Even though the nature of mirrors creates infinite number of imaginary receivers, molecules do not move as fast as photons. The first boundary condition, (5), implies that the  distribution of molecules vanishes at far greater distances. While $K$ is increasing, $\text{$r_{\text{Im$_{i}$}}$}$ increases for imaginary receivers at the edges. Therefore, an odd number could be selected for $K$, and then the analysis could be performed. Additionally, it could be assumed that for larger values of $d$, $K$ could be taken as a small number. The reason behind is that the distance between successive receivers is linearly proportional to $d+r_r$. When $d$ is large, (26) starts to saturate for small values of $K$ since $\text{$r_{\text{Im$_{i}$}}$}$ increases faster for large values of $d$. Therefore, after a certain small $K'$, the remaining summation becomes negligible. As a result, in exchange for a tolerable error, (26) could be simplified one step further for large values of $d$ by ignoring recursively effected molecules.
\begin{align}
    &p_{\text{hit}}^{\text{SISO}}(\text{Rx}, \text{$r_{\text{Im$_{0}$}}$}, r_r, t \mid \partial \Omega^{1},\partial \Omega^{2})  \nonumber \\ 
    &\approx \sum_{i=0}^{K'} \frac{r_r}{\text{$r_{\text{Im$_{i}$}}$}} \text{erfc}\left(\frac{\text{$r_{\text{Im$_{i}$}}$} - r_r}{\sqrt{4Dt}}\right) \nonumber \\
   & - \sum_{{j=0, j \ne i}}^{K'} \frac{r_r^2}{\text{$r_{\text{Im$_{j}$}}$} {r_0}_{\text{Rx$_{\text{Im$_i$}}$}|\text{Rx$_{\text{Im$_j$}}$}}} \nonumber \\ 
    &\times \text{erfc}\left(-\frac{(\text{$r_{\text{Im$_{i}$}}$} + {r_0}_{\text{Rx$_{\text{Im$_i$}}$}|\text{Rx$_{\text{Im$_j$}}$}}) - 2r_r}{\sqrt{4Dt}} \right).    \nonumber \\
\end{align}
The evaluation of (27) is shown in Fig. \ref{fig:double}, where $d_{min}(\partial \Omega ^{1},\text{Rx}) = d_{min}(\partial \Omega ^{2},\text{Rx}) = 3\mu m$.

\begin{table}[ht]
\renewcommand{\arraystretch}{1.1}
\caption{RMSE for Two Parallel Surface System}
\label{table_example}
\centering
\begin{tabular}{c||c||c}
\hline
\bfseries $K' = 3 $ & \bfseries $K' = 5$ & \bfseries $K' = 11 $ \\
\hline\hline
0.0089 & 0.0107 & 0.0106 \\
\hline
\end{tabular}
\end{table}

\subsection{Effect of $\sqrt{2D\Delta t}$ on RMSE Performances}

Performance analysis for the scenarios in which Rx is located very close to $\partial \Omega$ is vulnerable to the simulation parameter $\Delta t$. Therefore, $\Delta t$ must be chosen less than $10^{-5} s$ when $d_{min}$ is a very small number. The reason behind is that the motion of molecules, which are confined into a small region between Rx and $\partial \Omega$, is smoother at $\Delta t = 10^{-5} s$, giving better spacial resolution. In Table \ref{top2rms} and \ref{top3rms}, it appears that the RMSE tends to decrease with an increase in $d$. Furthermore, in Topology 0 and Topology 1, as $r_r$ increases, $d$ decreases as mentioned in Subsection A. Consequently, as observed in both Table III and Table V, the RMSE tends to increase with a decrease in $d$, consistent with our expectations.

\section{Conclusion}
In this paper, we propose deriving the MCvD channel response in a 3-D half-space with an infinite reflecting surface. It is demonstrated that an infinite reflecting surface creates a virtual channel behind it. Consequently, a SISO system in a 3-D half-space with an infinite reflecting surface can be approximated as a SIMO system in full 3-D space. This approximation obviates the need for the exhaustive and extremely difficult task of finding a closed-form solution by solving the diffusion equation with one initial and three boundary conditions. To validate the proposed method, various topologies are tested with results obtained from a GPU-based simulator. Moreover, the incremental step in Brownian motion becomes more significant in half-space simulations. Therefore, the appropriate choice of $\Delta t$ is crucial.

The results that are obtained through this work could play a crucial role in addressing localization problems within nanonetworks operating in biological environments.

\bibliographystyle{IEEEtran}
\bibliography{refs}

\begin{thebibliography}{10}
\providecommand{\url}[1]{#1}
\csname url@samestyle\endcsname
\providecommand{\newblock}{\relax}
\providecommand{\bibinfo}[2]{#2}
\providecommand{\BIBentrySTDinterwordspacing}{\spaceskip=0pt\relax}
\providecommand{\BIBentryALTinterwordstretchfactor}{4}
\providecommand{\BIBentryALTinterwordspacing}{\spaceskip=\fontdimen2\font plus
\BIBentryALTinterwordstretchfactor\fontdimen3\font minus \fontdimen4\font\relax}
\providecommand{\BIBforeignlanguage}[2]{{%
\expandafter\ifx\csname l@#1\endcsname\relax
\typeout{** WARNING: IEEEtran.bst: No hyphenation pattern has been}%
\typeout{** loaded for the language `#1'. Using the pattern for}%
\typeout{** the default language instead.}%
\else
\language=\csname l@#1\endcsname
\fi
#2}}
\providecommand{\BIBdecl}{\relax}
\BIBdecl

\bibitem{kuran2020survey}
M.~S. Kuran, H.~B. Yilmaz, I.~Demirkol, N.~Farsad, and A.~Goldsmith, ``A survey on modulation techniques in molecular communication via diffusion,'' \emph{IEEE Communications Surveys \& Tutorials}, vol.~23, no.~1, pp. 7--28, 2020.

\bibitem{gursoy2022towards}
M.~C. Gursoy, M.~Nasiri-Kenari, and U.~Mitra, ``Towards high data-rate diffusive molecular communications: A review on performance enhancement strategies,'' \emph{Digital Signal Processing}, vol. 124, p. 103161, 2022.

\bibitem{NAKANO}
T.~Nakano, M.~J. Moore, F.~Wei, A.~V. Vasilakos, and J.~Shuai, ``Molecular communication and networking: Opportunities and challenges,'' \emph{IEEE Transactions on NanoBioscience}, vol.~11, no.~2, pp. 135--148, 2012.

\bibitem{IAN}
\BIBentryALTinterwordspacing
I.~F. Akyildiz, J.~M. Jornet, and M.~Pierobon, ``Nanonetworks: A new frontier in communications,'' \emph{Commun. ACM}, vol.~54, no.~11, p. 84–89, nov 2011. [Online]. Available: \url{https://doi.org/10.1145/2018396.2018417}
\BIBentrySTDinterwordspacing

\bibitem{bi2021survey}
D.~Bi, A.~Almpanis, A.~Noel, Y.~Deng, and R.~Schober, ``A survey of molecular communication in cell biology: Establishing a new hierarchy for interdisciplinary applications,'' \emph{IEEE Communications Surveys \& Tutorials}, vol.~23, no.~3, pp. 1494--1545, 2021.

\bibitem{ningthoujam2021implementing}
S.~Ningthoujam and S.~K. Chakraborty, ``Implementing single path and multipath techniques under feedback channel for molecular communication,'' \emph{Wireless Personal Communications}, vol. 120, no.~4, pp. 3315--3328, 2021.

\bibitem{energy}
M.~Kuran, H.~B. Yilmaz, T.~Tugcu, and B.~Özerman Edis, ``Energy model for communication via diffusion in nanonetworks,'' \emph{Nano Communication Networks}, vol.~1, pp. 86--95, 06 2010.

\bibitem{chouhan2023interfacing}
L.~Chouhan and M.-S. Alouini, ``Interfacing of molecular communication system with various communication systems over internet of every nano things,'' \emph{IEEE Internet of Things Journal}, 2023.

\bibitem{receivedsignaling}
H.~B. Yilmaz and C.-B. Chae, ``Arrival modelling for molecular communication via diffusion,'' \emph{Electronics Letters}, vol.~50, pp. 1667--1669, 11 2014.

\bibitem{tunabirkan}
H.~B. Yilmaz, A.~C. Heren, T.~Tugcu, and C.-B. Chae, ``Three-dimensional channel characteristics for molecular communications with an absorbing receiver,'' \emph{IEEE Commun. Letters}, vol.~18, no.~6, pp. 929--932, 2014.

\bibitem{fatihdinc}
F.~Din{\c{c}}, B.~C. Akdeniz, A.~E. Pusane, and T.~Tugcu, ``Impulse response of the molecular diffusion channel with a spherical absorbing receiver and a spherical reflective boundary,'' \emph{IEEE Transactions on Molecular, Biological and Multi-Scale Communications}, vol.~4, no.~2, pp. 118--122, 2018.

\bibitem{absorbingspherical}
M.~M. Al-Zu’bi and A.~S. Mohan, ``Modeling of ligand-receptor protein interaction in biodegradable spherical bounded biological micro-environments,'' \emph{IEEE Access}, vol.~6, pp. 25\,007--25\,018, 2018.

\bibitem{synapticcell}
A.~E. Oncu, H.~U. Ozdemir, H.~I. Orhan, B.~C.~C. Akdeniz, A.~Toprakci, M.~A. Aslihak, H.~B. Yilmaz, A.~E. Pusane, T.~Tugcu, and F.~Dinc, ``Analytical investigation of long-time diffusion dynamics in a synaptic channel with glial cells,'' \emph{IEEE Communications Letters}, vol.~25, no.~11, pp. 3444--3448, 2021.

\bibitem{vessel}
M.~Turan, M.~S. Kuran, H.~B. Yilmaz, I.~Demirkol, and T.~Tugcu, ``Channel model of molecular communication via diffusion in a vessel-like environment considering a partially covering receiver,'' in \emph{2018 IEEE International Black Sea Conference on Communications and Networking (BlackSeaCom)}, 2018, pp. 1--5.

\bibitem{partially}
H.~Arjmandi, M.~Zoofaghari, and A.~Noel, ``Diffusive molecular communication in a biological spherical environment with partially absorbing boundary,'' \emph{IEEE Transactions on Communications}, vol.~67, no.~10, pp. 6858--6867, 2019.

\bibitem{chip}
N.~Farsad, A.~W. Eckford, S.~Hiyama, and Y.~Moritani, ``On-chip molecular communication: Analysis and design,'' \emph{IEEE Transactions on NanoBioscience}, vol.~11, no.~3, pp. 304--314, 2012.

\bibitem{bloodvess}
L.~Felicetti, M.~Femminella, and G.~Reali, ``Establishing digital molecular communications in blood vessels,'' in \emph{2013 First International Black Sea Conference on Communications and Networking (BlackSeaCom)}, 2013, pp. 54--58.

\bibitem{schulten2000lectures}
K.~Schulten and I.~Kosztin, ``Lectures in theoretical biophysics, vol. 117,'' \emph{Champaign, IL, USA: Univ. Illinois}, 2000.

\bibitem{Gokberk}
G.~Yaylali, B.~C. Akdeniz, T.~Tugcu, and A.~E. Pusane, ``Channel modeling for multi-receiver molecular communication systems,'' \emph{IEEE Transactions on Communications}, vol.~71, no.~8, pp. 4499--4512, 2023.

\end{thebibliography}

\end{document}